\documentclass[%
 reprint,
 amsmath,amssymb,
 aps,
]{revtex4-2}

\usepackage{graphicx}
\usepackage{dcolumn}
\usepackage{bm}
\usepackage[colorlinks,citecolor=blue]{hyperref} \usepackage{cleveref}
\hypersetup{colorlinks=true, linkcolor=blue, filecolor=magenta, urlcolor=blue,}
\usepackage[mathlines]{lineno}

\usepackage{subfigure}
\usepackage{xcolor}

\begin{document}

\preprint{APS/123-QED}

\title{Variational Monte Carlo Study of the Doped \emph{t}-\emph{J} Model on Honeycomb Lattice}

\author{Can Cui}
\author{Jing-Yu Zhao}
\author{Zheng-Yu Weng}%
\affiliation
{
Institute for Advanced Study, Tsinghua University, Beijing 100084, China }%

\date{\today}

\begin{abstract}
The ground state of the bipartite $t$-$J$ model must satisfy a specific sign structure, based on which the single-hole and two-hole ground state $Ans\ddot{a}tze$ on  honeycomb lattice are constructed and studied by a variational Monte Carlo (VMC) method. The VMC results are in good agreement with the exact diagonalization (ED) calculation. For the single-hole case, the degenerate ground states are characterized by quantum numbers of a spin-1/2 and an orbital angular momentum $L_z=\pm 2$. The latter is associated with the emergent chiral spin/hole currents mutually surrounding the hole/spin-1/2 as a composite object or ``twisted hole''. A vanishing quasiparticle spectral weight is shown in the large-sample limit. In the two-hole ground state, the holes form a spin-singlet pairing with $d$+$id$ symmetry in the Cooper channel, but are of $s$-wave symmetry as a tightly bound pair of the ``twisted holes''. Such a pairing mechanism of dichotomy can be attributed to eliminating the local spin currents which has nothing to do with the long-range antiferromagnetic correlation. Superconducting ground state at finite doping is briefly discussed in terms of the tightly bound hole pairs as the building blocks.
\end{abstract}

\maketitle


\section{Introduction}

Since the high-$T_c$ cuprate superconductor was discovered in 1986 \cite{Bednorz1986}, the microscopic mechanism behind it remains controversial despite more than three decades of intense investigation both experimentally and theoretically. Several basic facts related to the high-$T_c$ problem are as follows: (1) The ground state is an antiferromagnetic (AFM) long-range ordered state at half-filling; (2) Hole doping can quickly destroy the AFM order and gives rise to a superconducting (SC) state; (3) The pairing mechanism responsible for high-$T_c$ may be purely of an electronic origin, not mediated by vibration of lattices. 

The two-dimensional $t$-$J$ Hamiltonian has been proposed \cite{Anderson1987} as a minimal model to describe the physics in the cuprate as a doped Mott insulator. At half-filling, the $t$-$J$ model reduces to the Heisenberg model, whose ground state well describes the AFM state \cite{Liang1988}. Upon doping, a resonating valence bond (RVB) state \cite{Anderson1973,Anderson1987,RevModPhys.78.17} has been proposed as the parent state responsible for the pairing of the doped holes, which leads to the SC instability. However, how to connect the true AFM state at zero doping to a possible RVB state with strong pairing of the doped charges upon doping has been a great theoretical challenge \cite{RevModPhys.78.17}.

\begin{figure}[h]
\includegraphics[width=0.4\textwidth]{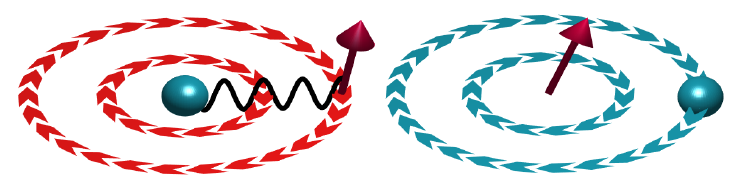}
\caption{\label{fig:mutual} Mutual entanglement between the hole and a spin-1/2 in the single-hole-doped ground state \emph{Ansatz} due to the phase-string sign structure. Here the hole is surrounded by the transverse spin current of the spin-1/2 and vice versa the latter is surrounded by the hole current, with the hole composite characterized by an emergent quantum number (angular momentum) $L_z=\pm 2$ on honeycomb lattice in contrast to $L_z=\pm 1$ on square lattice \cite{PhysRevB.99.205128}.}
\end{figure}

For an AFM Heisenberg model on a bipartite lattice, the ground state generally satisfies the so-called Marshall sign rule \cite{Marshall1955-gi}, which ensures the wave function to be spin singlet at any finite size. Once the holes are doped into such a spin background, it can be generally proved that the Marshall signs will be disordered by the hopping of the holes, leading to the so-called phase strings \cite{PhysRevLett.77.5102,Weng1997,Wu2008}.  The latter has two folds of non-perturbative consequences: One is that at each step of hole hopping, the propagator will be generally modulated by a singular sign $\pm$, depending on up- or down-spin exchanged with the hole; Second is that the accumulation of such signs over a closed path of each doped hole amounts to a topological Berry-like phase depending on the parity of the total down-spins (or up-spins due to the bipartite symmetry) exchanged with the hole. The former originates from the disordered Marshall signs which prevent a perturbative treatment of the doping problem even for the one-hole case, leading to an ``orthogonality catastrophe" effect \cite{PhysRevLett.18.1049,Anderson1990,PhysRevLett.77.5102}. As a generalized Berry-like phase, the latter 
implies a topological gauge structure associated with mutual ``semionic'' statistics between the spins and holes \cite{Weng1997,PhysRevLett.80.5401}. Here the mutual charge-spin entanglement due to the phase-string sign structure is schematically illustrated in Fig.~\ref{fig:mutual}.

The ground-state wave functions of the doped $t$-$J$ model have to generally satisfy the phase-string sign structure \cite{Wu2008} in replacing the Marshall sign upon doping. Demonstrating its non-perturbative nature, single-hole and two-hole ground states for the square lattice of the $t$-$J$ model have been recently investigated by exact diagonalization (ED), density matrix renormalization group (DMRG) numerical methods together with variational Monte Carlo (VMC) study based on variational $Ans\ddot{a}tze$ \cite{PhysRevB.98.165102,PhysRevB.99.205128,PhysRevX.12.011062}. The results have consistently revealed that the single-hole problem is indeed highly singular on the square lattice, and the previous perturbative approaches are problematic in producing the correct behavior. In particular, it is such a singularly frustrated motion of the single hole that leads to a strong local pairing between two doped holes \cite{PhysRevX.12.011062}, which becomes the building block for an unconventional pairing mechanism for the $t$-$J$ model at finite doping \cite{Weng_2011}. 

To further understand the novel propagations of a few doped holes in an AFM spin background, one may consider the $t$-$J$ model on a honeycomb lattice, in which the AFM order does not lead to a folding of the Brillouin zone in contrast to the square lattice, as each unit cell contains two sites with or without the AFM order. The honeycomb lattice is also a bipartite lattice where the Heisenberg Hamiltonian still satisfies the Marshall sign rule \cite{Marshall1955-gi}. Thus by doping, the same phase-string sign structure is present in the honeycomb lattice, whereas the effect of the long-range AFM may be distinct on the motion of doped holes. The latter has been argued in some analytical and numerical approaches investigating the doped  $t$-$J$ model or Hubbard model on honeycomb lattice recently \cite{Gu2013,Xu2023,PhysRevB.73.155118,PhysRevB.78.214406,PhysRevX.4.031040,miao2023spincharge,PhysRevB.90.054521,peng2023superconductivity,PhysRevB.105.035111,PhysRevB.97.075127}. 

In this paper, we construct the single-hole and two-hole ansatz states on the honeycomb lattice with properly incorporating the phase-string sign structure. Then we use the VMC method to explore the ground state properties in comparison with ED results on a small system as the benchmark. We shall identify the novel quantum numbers for the single-hole and two-hole ground states, which are fundamentally different from the construction based on the conventional Landau quasiparticles and their BCS-like Cooper pairing. For the former, the two-component structures have been revealed in both the single-hole and two-hole ground states. Here, even though the quasiparticle weight for the single hole and a $d+id$ pairing symmetry for the two holes can be identified, the dominant components are characterized by either vortex or the vortex-antivortex pair of spin currents surrounding the hole(s). The latter as the transverse phase-string effect cannot be deduced perturbatively. In particular, they lead to vanishing quasiparticle weight for the single hole in the large-sample limit and a strong local $s$-wave pairing for the two-hole composites insensitive to the long-range AFM order.

The rest of this paper is organized as follows: In Sec. \ref{sec:2}, we introduce the $t$-$J$ model and put forward a single-hole wave function \emph{Ansatz}. Based on the VMC study of the quantum numbers and ground-state energy, the spin-hole mutual entanglement pattern (cf. Fig.~\ref{fig:mutual}) is further revealed in terms of the hole and spin current patterns. The quasiparticle spectral weight is calculated which shows a non-Fermi-liquid behaviour in the large-sample size limit. By contrast, a Landau-quasiparticle-like behavior is recovered for the single hole in the $\sigma\cdot t$-$J$ model where the phase-string sign structure is precisely turned off. In Sec. \ref{sec:3}, we put forward a two-hole wave function \emph{Ansatz}. We will show the pairing symmetry in the Cooper channel is a $d+id$ wave, which emerges from the $s$ wave pairing symmetry of ``twisted quasiparticles''. The binding energy and pairing size of the hole pair are also calculated. Finally, we discuss the results and conclude in Sec. \ref{sec:4}.

\section{The single-hole ground state}\label{sec:2}

\subsection{The $t$-$J$ model}

\begin{figure}[h]
\centering
\includegraphics[width=0.25\textwidth]{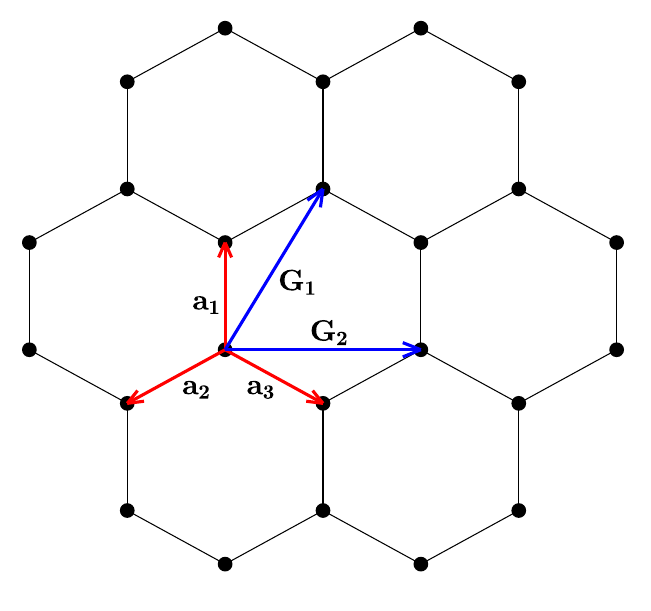}
\caption{\label{fig:Lattice_24} A 24-site honeycomb lattice with $\mathbf{a_{1}}$,$\mathbf{a_{2}}$,$\mathbf{a_{3}}$ denoting the bond vectors and $\mathbf{G_{1}}$,$\mathbf{G_{2}}$ two Bravais vectors. Here the lattice constant is set as $a_0=1$.
}
\end{figure}

In this paper, we shall study the variational ground states of the single-hole-doped and two-hole-doped $t$-$J$ model on a honeycomb lattice with open boundary condition (OBC) and $C_6$ rotation symmetry. 
A 24-site honeycomb lattice is illustrated in Fig.~\ref{fig:Lattice_24}. 

The Hamiltonian of the $t$-$J$ model is given by $H_{t\text{-}J}=\mathcal{P} (H_t + H_J) \mathcal{P}$, where hopping term $H_t$ and superexchange term $H_J$ are given by:
\begin{eqnarray}
H_{t}=-t\sum_{\langle ij\rangle,\sigma}(c_{i\sigma}^{\dagger}c_{j\sigma}+\mathrm{H.c.})~,\label{eq:Ht}
\\
H_J=J\sum_{\langle ij\rangle}\biggl(\mathbf{S}_i\cdot\mathbf{S}_j-\frac{1}{4}n_in_j\biggr)~.\label{eq:HJ}
\end{eqnarray}
Here $\mathcal{P}$ is the projection operator imposing the no-double-occupancy constraint: $\sum_{\sigma}c^{\dagger}_{i\sigma}c_{i\sigma}\leq 1$ on each site, which leads to the strong correlation of the electrons.
Throughout this paper, we choose the superexchange term $J=1$ as the unit, and fix the hopping term as $t/J=3$. 

Similar to the square lattice case, the hopping of the doped holes on the bipartite honeycomb lattice will disorder the Marshall signs hidden in the ground state of the half-filling to create the so-called phase string effect. By comparison, we shall also study the so-called $\sigma\cdot t$-$J$ model, which is similar to the above $t$-$J$ model but with the hopping term $H_t$ changed to:
\begin{equation}
H_{\sigma\cdot t}=-t\sum_{\langle ij\rangle,\sigma}\sigma c_{i\sigma}^{\dagger}c_{j\sigma}+\mathrm{H.c.},\label{eq:sigmat_J}
\end{equation}
where an additional spin-dependent sign $\sigma=\pm 1$ is inserted. It can be straightforwardly shown that in the single-hole doped $\sigma\cdot t$-$J$ model,   
the off-diagonal matrix elements for both $H_{\sigma\cdot t}$ and $H_J$ can be made negative semidefinite in the hole and spin Ising basis (e.g., defined with $S^z$ as the quantization axis) if the Marshall signs are explicitly incorporated. It means that the ground state must satisfy the Marshall sign without significant frustration introduced by hole doping. This is in sharp contrast to the $t$-$J$ model where the hopping term will explicitly violate the Marshall sign rule to result in the phase-string sign structure. A comparative study of the two models can further reveal the nature of strong correlation in the $t$-$J$ model.

\subsection{Wave function \emph{Ansatz} of the single-hole ground state}

In the following, we shall study a single-hole wave function \emph{Ansatz}:
\begin{equation}
|\Psi_{m}\rangle_{1\mathrm{h}}=\sum_{i}\varphi_{\mathrm{h}}^{(m)}(i)c_{i\uparrow}e^{-im\hat{\Omega}_{i}}|\phi_{0}\rangle, \label{eq:singlehole}
\end{equation}
in which a ``spin-up'' electron is moved from the half-filling spin background $|\phi_0 \rangle$. Here Eq.~(\ref{eq:singlehole}) is of the same form as that previously proposed for the square lattice case \cite{PhysRevB.99.205128}, with $|\phi_0 \rangle$ representing the ground state at half filling, i.e. that of the Heisenberg Hamiltonian in Eq.~(\ref{eq:HJ}). Such $|\phi_0 \rangle$ is well simulated by using the Liang-Doucot-Anderson-type variational wave function \cite{Liang1988} on the honeycomb lattice (cf. Appendix~\ref{app:secA}). Then, the single-hole variational wave function $\varphi_{\mathrm{h}}^{(m)}(i)$ in Eq.~(\ref{eq:singlehole}) can be further determined by VMC. 

If one takes $m=0$ in Eq.~(\ref{eq:singlehole}), the ansatz state reduces to a Bloch-wave-like state for the doped hole:
\begin{equation}
|\Psi_{0}\rangle_{1\mathrm{h}}=\sum_{i}\varphi_{\mathrm{h}}^{\mathrm{B}}(i)c_{i\uparrow}|\phi_{0}\rangle, \label{eq:Bloch}
\end{equation}
with $\varphi_{\mathrm{h}}^{\mathrm{B}}(i)\propto e^{i\mathbf{k}\cdot\mathbf{r}_i}$ for a translationally invariant system, where $\mathbf{k}$ is the wave vector of the Bloch electron (hole). This wave function in Eq.~(\ref{eq:Bloch}) will satisfy the Marshall sign structure of $\sigma\cdot t$-$J$ model as obeyed by $|\phi_0 \rangle$, such that it may serve as a variational wave function for the $\sigma\cdot t$-$J$ model in Eq.~(\ref{eq:sigmat_J}) (with a trivial $\mathbf{k}=0$ in the ground state). 

At $m=\pm 1$, Eq.~(\ref{eq:singlehole}) will differ from the usual Bloch-wave-like quasiparticle state by the many-body phase shift $\hat{\Omega}_{i}$, which is given by \cite{Weng_2011,Weng2011}
\begin{equation}
\hat{\Omega}_i=\sum_{l(\neq i)}\theta_i(l)n_{l\downarrow},\label{eq:Omega}
\end{equation}
where $n_{l\downarrow}=c_{l\downarrow}^{\dagger}c_{l\downarrow}$ measures the number of the spin-down electron at site $l$ and $\theta_{i}(l)=\mathrm{Im}\ln(z_{i}-z_{l})$ is the statistical angle expanded between a straight line connecting site $i$ and $l$, and the horizontal axis (with $z_{i}=x_i+iy_i$ denoting the complex coordinate of site $i$). 
Here such a many-body phase shift operator as induced by the doped hole can contribute to a sign change each time when an exchange of the doped hole at site $i$ with a $\downarrow$ spin at the nearest-neighbor site $j$ occurs due to $\theta_i(j)-\theta_j(i)=\pm \pi$, which precisely compensates the sequence of signs that the hopping of the hole must pick up in the $t$-$J$ model, i.e., the phase-string effect $(+1)\times (-1)\times (+1) ... $ depending on the spins exchanged with the hole. Of course, the other sites in the summation of Eq. (\ref{eq:Omega}) will not be completely compensated and contribute to both the hopping and superexchange terms by some topological link variables known as the mutual Chern-Simon gauge fields which are much more smoothed and perturbatively treatable \cite{Weng_2011}. Indeed, as a unitary transformation, $e^{\mp i\hat{\Omega}_{i}}$ here should not change the physical consequence of the phase string as a generalized Berry phase for each closed loop, which is to be captured precisely by the mutual Chern-Simons gauge fields after the unitary transformation. However, it significantly \emph{regulates} the hopping term by replacing the singular $\pm $ signs with the weakly fluctuating mutual Chern-Simons link variables \cite{Weng_2011}. 

In the present VMC approach, the single-hole wave function $\varphi_{\mathrm{h}}^{(m)}(i)$ as a smooth functional after the ``duality transformation'' in Eq.~(\ref{eq:singlehole}) can be decided variationally. The ground state energy and the quantum number obtained by VMC and ED calculations for the 24-site system (cf. Fig.~\ref{fig:Lattice_24}) are listed in Table~\ref{tab:table1} for comparison.

\begin{table}[h]
\caption{\label{tab:table1}%
The single-hole ground state energies and quantum numbers in the 24-site system are calculated by VMC and ED methods. $E_{tot}$ denotes the total energy, while $E_t$ and $E_J$ are the kinetic and superexchange energies, respectively. $L_z$ is the angular momentum defined by $\hat{R}(\theta)|\Psi\rangle=e^{iL_z\theta}|\Psi\rangle$, where $\hat{R}(\theta)$ is the generator of rotation in the $C_6$ group.
}
\begin{ruledtabular}
\begin{tabular}{ccccc}
&
\textrm{$E_{tot}$}&
\textrm{$E_t$}&
\textrm{$E_J$}&
\textrm{$L_z$}\\ 
\colrule
$|\Psi_{0}\rangle_{1\mathrm{h}}$ & -20.189 & -2.224 & -17.965 & 0\\
$|\Psi_{\pm 1}\rangle_{1\mathrm{h}}$ & -22.934 & -5.588 & -17.346 & $\pm 2$\\
$|\tilde{\Psi}_{\pm 1}\rangle_{1\mathrm{h}}$ & -24.241 & -7.032 & -17.209 & $\pm 2$\\
ED & -24.782 & -7.553 & -17.224 & $\pm 2$\\
\end{tabular}
\end{ruledtabular}
\end{table}


The ED result of Table~\ref{tab:table1} shows that the ground states of this model are double degenerate with $L_z=\pm 2$ for each given $S^z=\pm 1/2$. The VMC shows that $|\Psi_{\pm 1}\rangle_{1\mathrm{h}}$ can capture this quantum number precisely: $L_z=+2/-2$ corresponds to the choice of $m=+1/-1$, respectively. By contrast, the Bloch-wave-like state $|\Psi_{0}\rangle_{1\mathrm{h}}$ gives rise to a wrong $L_z=0$ without incorporating the phase string sign structure. 

With $e^{\pm i\hat{\Omega}_{i}}$, the variational ground-state energy $E_{tot}$ of $|\Psi_{\pm 1}\rangle_{1\mathrm{h}}$ can be improved significantly compared to the Bloch-wave-like state $|\Psi_{0}\rangle_{1\mathrm{h}}$. To be self-consistent, one may further incorporate a ``longitudinal spin polaron effect" to the spin background:  $|\phi_0\rangle \rightarrow \hat{\Pi}_i|\phi_0\rangle$ as
\begin{equation}
|\tilde{\Psi}_{m}\rangle_{1\mathrm{h}}=\sum_{i}\varphi_{\mathrm{h}}^{(m)}(i)c_{i\uparrow}e^{-im\hat{\Omega}_{i}}\hat{\Pi}_{i}|\phi_{0}\rangle, \label{eq:tildePsi}
\end{equation}
in which the detailed operator $\Pi_{i}$ is presented in Appendix~\ref{app:secD}. Consequently the variational energy of $|\tilde{\Psi}_{\pm 1}\rangle_{1\mathrm{h}}$ can be further improved within 2.18\% as compared to the exact ED result with the quantum number $L_z=\pm 2$ unchanged [see Table~\ref{tab:table1}].

\subsection{Hidden spin current related to $L_z=\pm 2$}

As shown in Table~\ref{tab:table1}, the single-hole ground state with a total $S^z=-1/2$ is degenerate with nontrivial angular momentum $L_z=\pm 2$. In the following, we shall identify a spin current surrounding the hole and \emph{vice versa} a hole current around the spin $S^z=-1/2$, as schematically illustrated in Fig.~\ref{fig:mutual}, which leads to the nontrivial $L_z$. In particular, the chirality of the spin/hole current is associated with the sign of $m=\pm 1$ in Eq.~(\ref{eq:singlehole}) and Eq.~(\ref{eq:tildePsi}). 

In general, the hole and spin currents can be deduced from the continuity equations:
\begin{eqnarray}
J_{i j}^h&=&i t \sum_\sigma\left(c_{i \sigma}^{\dagger} c_{j \sigma}-c_{j \sigma}^{\dagger} c_{i \sigma}\right),\label{eq:J_h}\\
J_{i j}^s&=&\frac{J}{2} i\left(S_i^{+} S_j^{-}-S_i^{-} S_j^{+}\right) ,\label{eq:J_s}\\
J_{i j}^b&=&-i \frac{t}{2} \sum_\sigma \sigma\left(c_{i \sigma}^{\dagger} c_{j \sigma}-c_{j \sigma}^{\dagger} c_{i \sigma}\right),\label{eq:J_b}
\end{eqnarray}
where $J^{h}$ is the hole current, and the spin current contains two parts: $J^{s}$ is the neutral spin current originating from spin exchange interaction and $J^{b}$ is the backflow spin current caused by the motion of the hole.

\begin{figure}[h]
\includegraphics[width=0.47\textwidth]{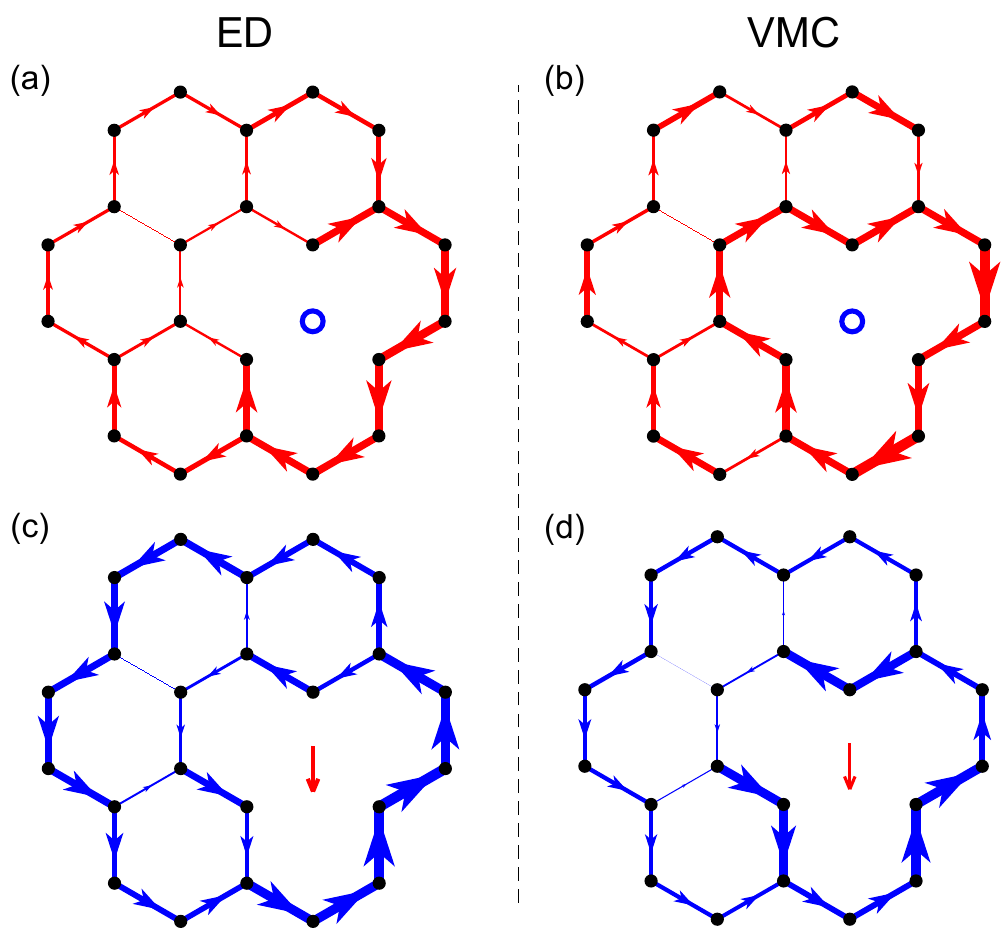}
\caption{\label{fig:Js_and_Jh}(a) and (b): ED and VMC calculations of $\langle \mathcal{P}_{l}^{h}J_{ij}^{s}\rangle$, respectively; (c) and (d): ED and VMC calculations of $\langle \mathcal{P}_{l}^{s}J_{ij}^{h}\rangle$, respectively. Here the ground states with $L_z=-2$ are examined, and the thickness of the bonds represents the relative magnitudes of the spin and hole currents in (a)-(d). }
\end{figure}

Here we use the variational wave function $|\tilde{\Psi}_{-1}\rangle_{1\mathrm{h}}$ [Eq.~(\ref{eq:tildePsi})] to examine the spin and hole currents. To demonstrate the entanglement between spin and hole, we calculate the spin current/hole current surrounding the hole/spin $\downarrow$ projected at a given site. The correlators $\langle \mathcal{P}_{l}^{h}J_{ij}^{s}\rangle$ and $\langle \mathcal{P}_{l}^{s}J_{ij}^{h}\rangle$ calculated by ED and VMC are presented in Fig.~\ref{fig:Js_and_Jh} for comparison, where $\mathcal{P}_{l}^{h}\equiv n_{l}^{h}=1-\sum_{\sigma}c^{\dagger}_{l\sigma}c_{l\sigma}$ and $\mathcal{P}_{l}^{s}\equiv n_{l\downarrow}$ project the hole and a $\downarrow$ spin at site $l$, respectively.

Figure ~\ref{fig:Js_and_Jh} shows both ED and VMC results that clearly illustrate the mutual entanglement pattern between the hole and the $\downarrow$ spin (cf. Fig.~\ref{fig:mutual}), which originates from the many-body phase shift operator $e^{- im\hat{\Omega}_{i}}$ in the variational wave function. Here the transverse spin current pattern around the hole is not present in the Bloch-like wave function, which reveals a non-Fermi liquid behavior in the one-hole limit. The chirality of the spin/hole currents will change sign when a variational state $|\Psi_{+1}\rangle_{1\mathrm{h}}$ is used with $L_z=+2$. 
If we further consider the spin degree of freedom associated with spin-1/2 introduced by the doped hole, the vortex patterns of the hole current and neutral spin current for the four degenerate states in the OBC system with $C_6$ symmetry are listed in Table.~\ref{tab:chirality}.

\begin{table}[h]
\caption{\label{tab:chirality}%
The single-hole quantum numbers, and the chiralities of the hole current $J^h$ and the neutral spin current $J^s$, which are associated with the phase-shift operators $\pm \hat{\Omega}_{i}$ in the single-hole variational wave function. }

\begin{ruledtabular}
\begin{tabular}{ccccc}
phase factor & $e^{i\hat{\Omega}_{i}}$ & $e^{-i\hat{\Omega}_{i}}$ & $e^{i\hat{\Omega}_{i}}$ &$e^{-i\hat{\Omega}_{i}}$\\
$S^z$ & $1/2$ & $1/2$ & $-1/2$ &$-1/2$\\
$L_z$ & $+2$ & $-2$ & $-2$ &$+2$\\
$J^h$ & $\curvearrowright$ & $\curvearrowleft$ & $\curvearrowleft$ & $\curvearrowright$\\
$J^s$ & $\curvearrowright$ & $\curvearrowleft$ & $\curvearrowright$ &$\curvearrowleft$\\
\end{tabular}
\end{ruledtabular}
\end{table}

\subsection{Quasiparticle spectral weight $Z_{\mathbf{k}}$ and non-Fermi-liquid behavior}


In the following, we examine the quasiparticle spectral weight function of the single-hole ground state by ED and VMC. Note that for the honeycomb lattice,  there are two sites per unit cell. A free-electron tight-binding model based on Eq.~(\ref{eq:Ht}) (without the no-double-occupancy projection) is given as follows
\begin{equation}
\begin{aligned}
H_0 & \equiv \sum_{\mathbf{k} \sigma}\left(\begin{array}{ll}
c_{A\mathbf{k}\sigma}^{\dagger} & c_{B\mathbf{k}\sigma}^{\dagger}
\end{array}\right)\left(\begin{array}{cc}
0 & -t f(\mathbf{k}) \\
-t f^*(\mathbf{k}) & 0
\end{array}\right)\left(\begin{array}{l}
c_{A\mathbf{k}\sigma} \\
c_{B\mathbf{k}\sigma}
\end{array}\right) \\
& =\sum_{\mathbf{k} \sigma}\left(\begin{array}{cc}
\alpha_{\mathbf{k}\sigma}^{\dagger} & \beta_{\mathbf{k}\sigma}^{\dagger}
\end{array}\right)\left(\begin{array}{cc}
t|f(\mathbf{k})| & 0 \\
0 & -t\left|f(\mathbf{k})\right|
\end{array}\right)\left(\begin{array}{c}
\alpha_{\mathbf{k}\sigma} \\
\beta_{\mathbf{k}\sigma}
\end{array}\right),
\end{aligned}
\end{equation}
where $f(\mathbf{k})=\sum\limits_{j=1}\limits^{3}e^{i\mathbf{k}\cdot\mathbf{a_j}}$, $c_{A\mathbf{k}\sigma}$ and $c_{B\mathbf{k}\sigma}$ denotes the electron operators on sublattice A and B, and $\mathbf{a_j}$ are defined in Fig.~\ref{fig:Lattice_24}. Then the quasiparticle spectral weight is defined by:
\begin{eqnarray}
Z_{\mathbf{k} 1} &\equiv&\left|\left\langle\operatorname{\phi_0}\left|\alpha_{\mathbf{k} \uparrow}^{\dagger}\right| \Psi_{G}\right\rangle_{1\mathrm{h}}\right|^2,\label{eq:Zkupper}\\
Z_{\mathbf{k} 2} &\equiv&\left|\left\langle\operatorname{\phi_0}\left|\beta_{\mathbf{k} \uparrow}^{\dagger}\right| \Psi_{G}\right\rangle_{1\mathrm{h}}\right|^2,\label{eq:Zklower}
\end{eqnarray}
where $Z_{\mathbf{k} 1}$ and $Z_{\mathbf{k} 2}$ denote two bands described by $\alpha_{\mathbf{k}}$ and $\beta_{\mathbf{k}}$ respectively.  

\begin{figure}[h]
\includegraphics[width=0.45\textwidth]{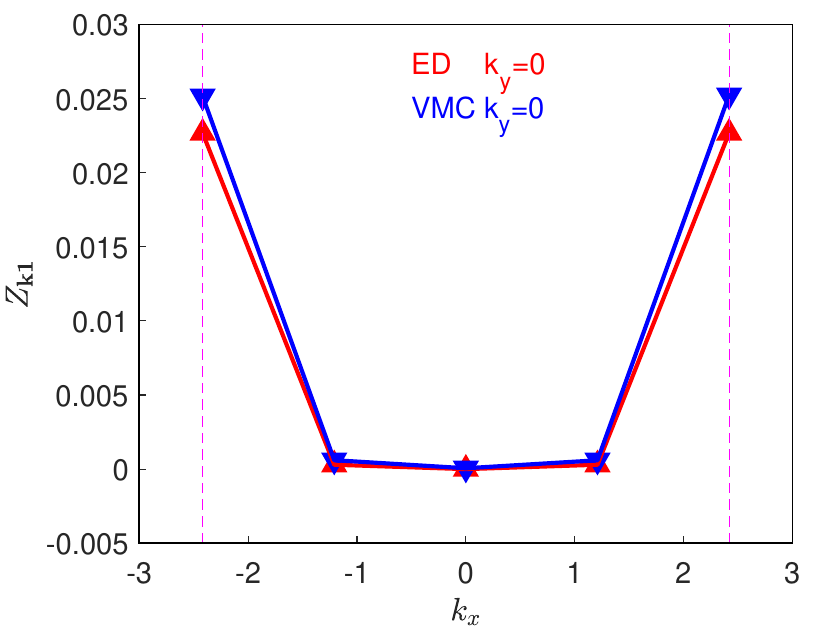}
\caption{\label{fig:ZkXaxis} $Z_{\mathbf{k}1}$ calculated by VMC and ED along $k_x$ direction in the 24-site system. The dashed lines mark the Brillouin zone boundary. }
\end{figure}

\begin{figure}[h]
\includegraphics[width=0.45\textwidth]{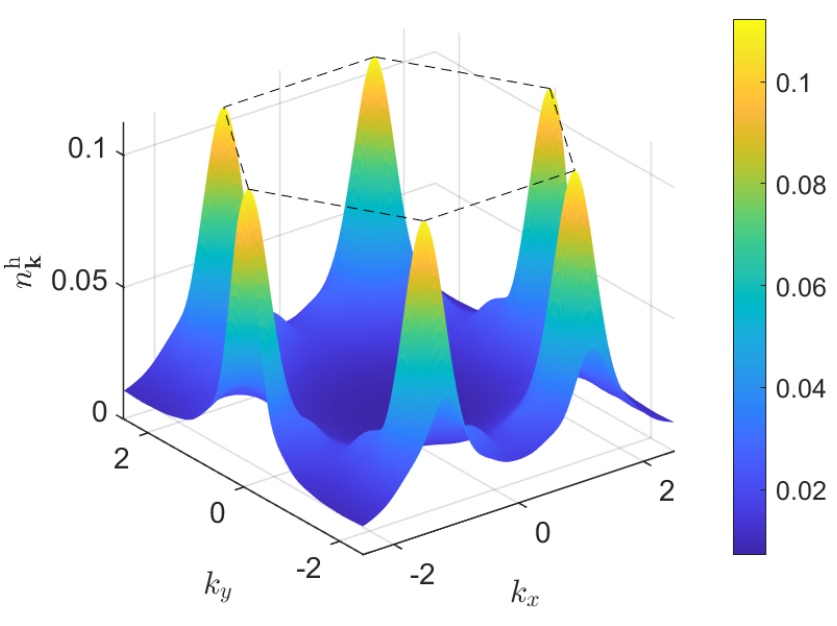}
\caption{\label{fig:nkh_96} The hole momentum distribution function, $n^{\mathrm{h}}_{\mathbf{k}}$, based on the ansatz state $|\Psi_{-1}\rangle_{1\mathrm{h}}$ on the 96-site honeycomb lattice, which peaks at the $K$ and $K'$ points with a broad distribution in the whole Brillouin zone. }
\end{figure}

The calculated $Z_{\mathbf{k}1}$ along $k_x$ direction in the 24-site honeycomb system [cf. Fig.~\ref{fig:Lattice_24}] is shown in Fig.~\ref{fig:ZkXaxis} in which the ED and VMC (using $|\tilde{\Psi}_{-1}\rangle_{1\mathrm{h}}$) are in good agreement ($Z_{\mathbf{k}2}$ is the same and not shown in the figure). The spectral weight $Z_{\mathbf{k}1}$ shows two independent peaks located at the vertices of the magnetic Brillouin zone boundary, which coincides with the peaks of the momentum distribution of the doped hole defined by
\begin{equation}
    n^{\mathrm{h}}_{\mathbf{k}}\equiv 2-\sum_{\sigma}(\alpha^{\dagger}_{\mathbf{k}\sigma}\alpha_{\mathbf{k}\sigma}+\beta^{\dagger}_{\mathbf{k}\sigma}\beta_{\mathbf{k}\sigma}), \label{eq:nkhole}
\end{equation}
which is presented in Fig.~\ref{fig:nkh_96} for the VMC calculation in the 96-site honeycomb lattice based on $|\Psi_{-1}\rangle_{1\mathrm{h}}$.

\begin{figure}[h]
\includegraphics[width=0.40\textwidth]{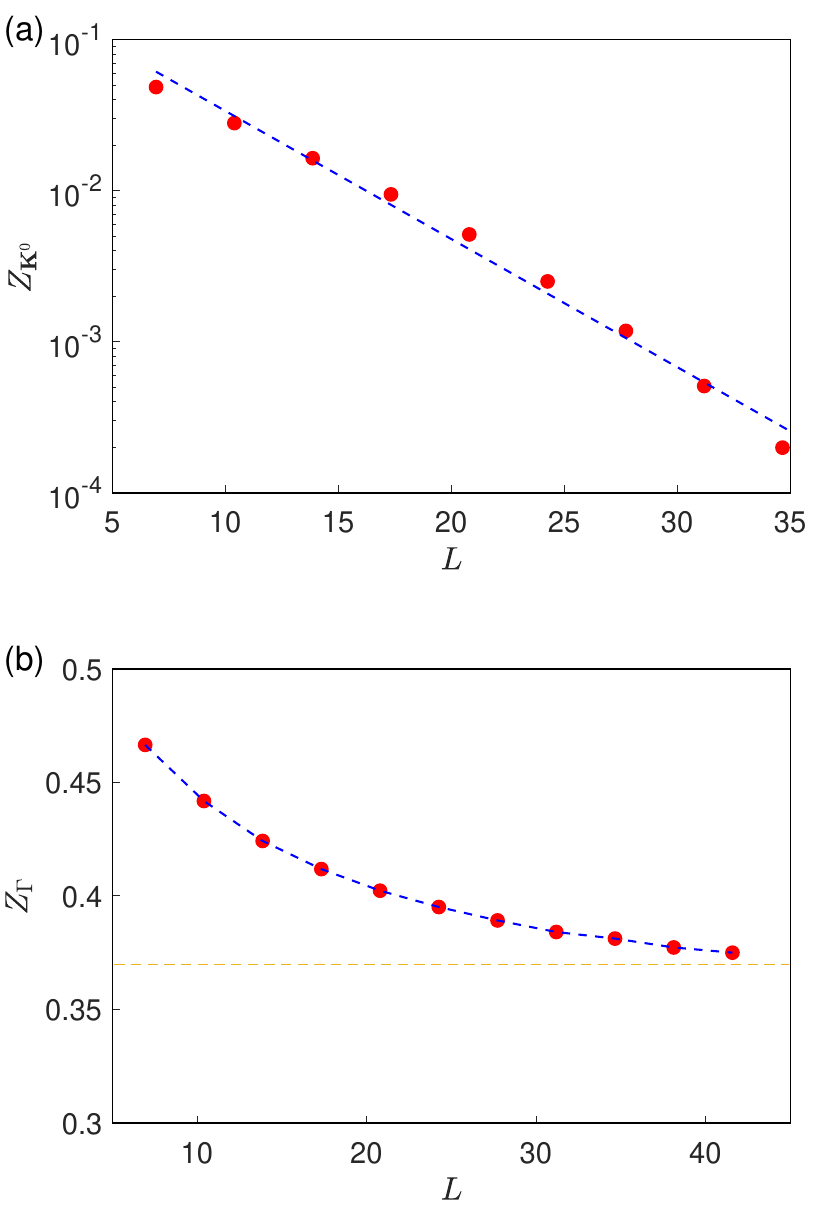}
\caption{\label{fig:Zk_tj_sigmatj}(a) The spectral weight $Z_{\mathbf{K}^0}$ vanishes in the large sample-length $L$ along the $\hat{x}$ direction. Calculated by VMC for the $t$-$J$ model, the slope of $\ln Z_{\mathbf{K}^0} $ is about $-1/5.119$. (b) In $\sigma\cdot t$-$J$ model, the spectral weight $Z_{\Gamma}$ at $\Gamma$ point versus $L$ is scaled to a finite value.}
\end{figure}

The momentum distribution in Fig.~\ref{fig:nkh_96} indicates a two-component structure in the ground-state wave function: besides the coherent Bloch-wave-like component with six peaks proportional to $Z_{\mathbf{k}}$, there is an incoherent component corresponding to a broad distribution in $n^{\mathrm{h}}_{\mathbf{k}}$, which can be attributed to the momentum transfer to the transverse spin current pattern surrounding the doped hole (cf. Fig.~\ref{fig:Js_and_Jh}). To show the importance of the incoherent component, the coherent component as represented by the quasiparticle spectral weight $Z_{\mathbf{k}1}$ as a function of the sample size is shown in Fig.~\ref{fig:Zk_tj_sigmatj} for both the $t$-$J$ model and $\sigma\cdot t$-$J$ models, respectively. In Fig.~\ref{fig:Zk_tj_sigmatj}(a), $Z_{\mathbf{K}^0}$ denotes the peak value for the $t$-$J$ model at $\mathbf{K}^0=(\frac{4\sqrt{3}\pi}{9},0)$. The scaling analysis of Fig.~\ref{fig:Zk_tj_sigmatj}(a) shows an exponential decay behaviour:
\begin{equation}
Z_{\mathbf{K}^0} \simeq 0.238 \,e^{-L/5.119}, \label{eq:Zkscaling}
\end{equation}
which reveals that the doped hole behaves as a non-Landau quasiparticle. By contrast, for the $\sigma\cdot t$-$J$ model, $Z_{\mathbf{k}1}$ is peaked at a unique $\Gamma=(0,0)$ denoted by $Z_{\Gamma}$ in Fig.~\ref{fig:Zk_tj_sigmatj}(b). As calculated by VMC, its variational state in Eq.~(\ref{eq:Bloch}) is non-degenerate with $L_z=+3$ in consistency with the ED calculation, where $Z_{\Gamma}$ converges to a finite value in the large system size limit (cf. Fig.~\ref{fig:Zk_tj_sigmatj}(b)). With ``turning off'' the phase string effect in the $\sigma\cdot t$-$J$ model, a finite $Z_{\Gamma}$ suggests the Bloch-wave-like component of the wave function remains robust in the thermodynamic limit.

\section{The two-hole ground state}\label{sec:3}
\subsection{Wave function \emph{Ansatz} and basic properties of the two-hole ground state}

Based on the single-hole wave function \emph{Ansatz} given in Eq.~(\ref{eq:singlehole}), a two-hole wave function \emph{Ansatz} may be straightforwardly constructed:
\begin{equation}
|\Psi_{m}\rangle_{2\mathrm{h}}=\sum_{i j} g_m(i, j) c_{i \uparrow} c_{j \downarrow} e^{+im( \hat{\Omega}_i- \hat{\Omega}_j)}|\phi_0\rangle,\label{eq:twohole}
\end{equation}
where the pairing wave function $g_m(i, j)$ is to be determined by VMC.

It can be checked that the time reversal operator $\hat{T}$ will change one type of the ansatz state to the other, i.e. $\hat{T}|\Psi_{m}\rangle_{2\mathrm{h}}=e^{i\theta_{T}}|\Psi_{-m}\rangle_{2\mathrm{h}}$, where $\theta_T$ is an arbitrary phase factor. Under the spin-flip transformation $\hat{F}$, the two wave functions remain unchanged with an arbitrary $U(1)$ phase factor $\theta_{F}$: $\hat{F}|\Psi_{m}\rangle_{2\mathrm{h}}=e^{i\theta_{F}}|\Psi_{m}\rangle_{2\mathrm{h}}$. The details are shown in Appendix~\ref{app:secB}.

\begin{figure*}[t]
\includegraphics[width=0.8\textwidth]{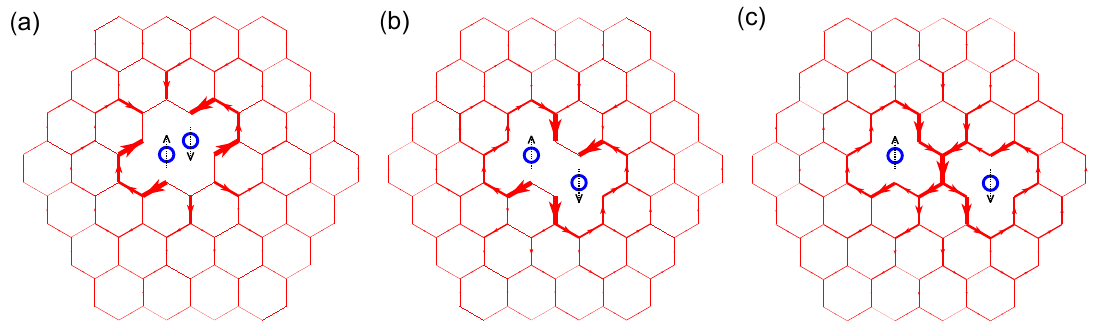}
\caption{\label{fig:Js_2h} The neutral spin current pattern emerges with two holes projected at fixed locations in the two-hole ansatz state $|\Psi_{+1}\rangle_{2\mathrm{h}}$. Blue circles with the dashed arrows represent the projected hole with spin $\uparrow$ or $\downarrow$. As the distance between the two holes increases in (a)-(c), the vortex-antivortex pair of the surrounding spin currents gets larger and stronger in strength.}
\end{figure*}

The ground state energy and angular momentum calculated by VMC and ED are listed in Table~\ref{tab:table2} for the 24-site system (cf. Fig~\ref{fig:Lattice_24}). In Table~\ref{tab:table2}, the Bloch-wave-like state $|\Psi_{0}\rangle_{2\mathrm{h}}$ with $m=0$ gives rise to an incorrect $L_z$ in comparison with the ED. The quantum number of the states $|\Psi_{\pm 1}\rangle_{2\mathrm{h}}$ are found to be $L_z=\pm 2$, respectively, in agreement with the ED result with a good VMC energy. Note that the energy of $|\Psi_{\pm 1}\rangle_{2\mathrm{h}}$ can be further improved by considering the ``longitudinal spin polaron effect" as in the single-hole case. 

\begin{table}[h]
\caption{\label{tab:table2}%
The two-hole ground state energies and quantum numbers in the 24-site system as calculated by VMC and ED methods. Here $E_{tot}$ is the total energy, $E_t$ and $E_J$ are the kinetic and superexchange energies. respectively.  $L_z$ denotes the angular momentum with total spin $S=0$.
}
\begin{ruledtabular}
\begin{tabular}{ccccc}
\textrm{}&
\textrm{$E_{tot}$}&
\textrm{$E_t$}&
\textrm{$E_J$}&
\textrm{$L_z$}\\
\colrule
$|\Psi_{0}\rangle_{2\mathrm{h}}$ & -21.498 & -5.000 & -16.498 & $0$\\
$|\Psi_{\pm 1}\rangle_{2\mathrm{h}}$ & -26.676 & -10.909 & -15.767 & $\pm 2$\\
ED & -29.834 & -14.433 & -15.401 & $\pm 2$\\
\end{tabular}
\end{ruledtabular}
\end{table}  

Note that null spin current is present in the ground state on account of the spin-flip symmetry mentioned above. But one may find an emergent neutral spin current pattern $\langle J^s_{ij} \rangle$ with two holes projected at some fixed sites as shown in Fig.~\ref{fig:Js_2h}. In the single-hole case, we have already seen that the hole is always surrounded by the spin currents, originated from the many-body phase shift $e^{\pm i\hat{\Omega}_i}$ (cf. Fig.~\ref{fig:Js_and_Jh}). The two-hole wave function in Eq.~(\ref{eq:twohole}) indicates that the spin currents surrounding the two doped holes are opposite in chiralities as shown in Figs.~\ref{fig:Js_2h}(a)-(c), which illustrates that as the two holes approach closer, the vortex and antivortex pair of the spin currents will come to shrink with diminishing net spin currents. Namely the pairing force for the two holes to form a tightly bound pair can be identified as the tendency to eliminate the spin currents associated with the hopping-induced phase string effect in the wave function. By contrast, the pairing is negligible in $|\Psi_{0}\rangle_{2\mathrm{h}}$ with the absence of $e^{\pm i\hat{\Omega}_i}$ in the latter. 

\begin{figure}[h]
\includegraphics[width=0.45\textwidth]{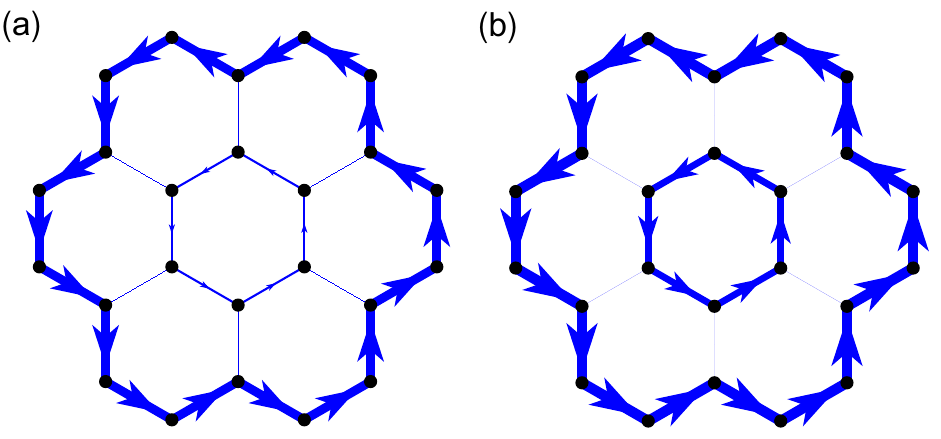}
\caption{\label{fig:Jh_2h} The chiral charge currents in the two-hole ground state with $L_z=+2$ as calculated by  (a) ED and (b) VMC for the 24-site lattice. The thickness of the bonds represents the relative magnitude of the hole current. }
\end{figure}

With the spin current being canceled on average in the two-hole state, a finite angular momentum $L_z=\pm 2$ of the degenerate ground state is then solely associated with a net chiral hole-current pattern. The hole-current distributions of the $L_z=+2$ state as calculated by ED and VMC, are shown in Figs.~\ref{fig:Jh_2h}(a) and \ref{fig:Jh_2h}(b), respectively. The chiral hole currents in the $L_z=-2$ state are opposite to those in Fig.~\ref{fig:Jh_2h}.

\subsection{$d+id$ symmetry of the Cooper pairing}

In the following, we show that the quantum number $L_z=\pm 2$ in the two-hole ground states will correspond to a $d+id$ symmetry of the Cooper pairing. Here the real space Cooper pair order parameter in the spin-singlet channel is defined by
\begin{equation}
\Delta_{i j}= _{2\mathrm{h}}\langle\Psi_m\left|c_{i \uparrow} c_{j \downarrow}-c_{j \uparrow} c_{i \downarrow}\right|\phi_0\rangle. \label{eq:Delta_ij}
\end{equation}

Denote $\Delta_{ij}=\Delta_{\mathbf{b}}$, where $\mathbf{b}=\mathbf{r}_j-\mathbf{r}_i$. Let us fix the site $i$ near the center of a 150-site sample and calculate the pairing order parameter of the nearest neighbors, i.e. $\mathbf{b}=\mathbf{a_1}, \mathbf{a_2}, \mathbf{a_3}$ in Fig.~\ref{fig:Lattice_24} based on $|\Psi_{2\mathrm{h}}\rangle=|\Psi_{+1}\rangle_{2\mathrm{h}}$ with $L_z=+2$. The ratios between $\Delta_{\mathbf{a_{i}}}$ are listed below:
\begin{eqnarray}
& \frac{\Delta_{\mathbf{a_2}}}{\Delta_{\mathbf{a_1}}}=-0.4961-0.8628 i, \nonumber \\
& \frac{\Delta_{\mathbf{a_3}}}{\Delta_{\mathbf{a_2}}}=-0.5045-0.8691 i,  \label{eq:Delta_ij_results}\\
& \frac{\Delta_{\mathbf{a_1}}}{\Delta_{\mathbf{a_3}}}=-0.4994-0.8663 i. \nonumber
\end{eqnarray} 
Here all of the ratios are close to $e^{i\frac{4\pi}{3}}$, indicating a $d+id$ pairing symmetry. Note that the small deviation from the exact $e^{i\frac{4\pi}{3}}$ comes from the deviation from the exact $C_3$ symmetry for the site $i$ in such a finite-size sample. And $|\Psi_{-1}\rangle_{2\mathrm{h}}$ with $L_z=-2$ corresponds to a $d-id$ pairing symmetry. The $d+id$ pairing symmetry is also identified by the tensor network method at finite but small doping limit \cite{Gu2013}.

\begin{figure}[h]
\includegraphics[width=0.4\textwidth]{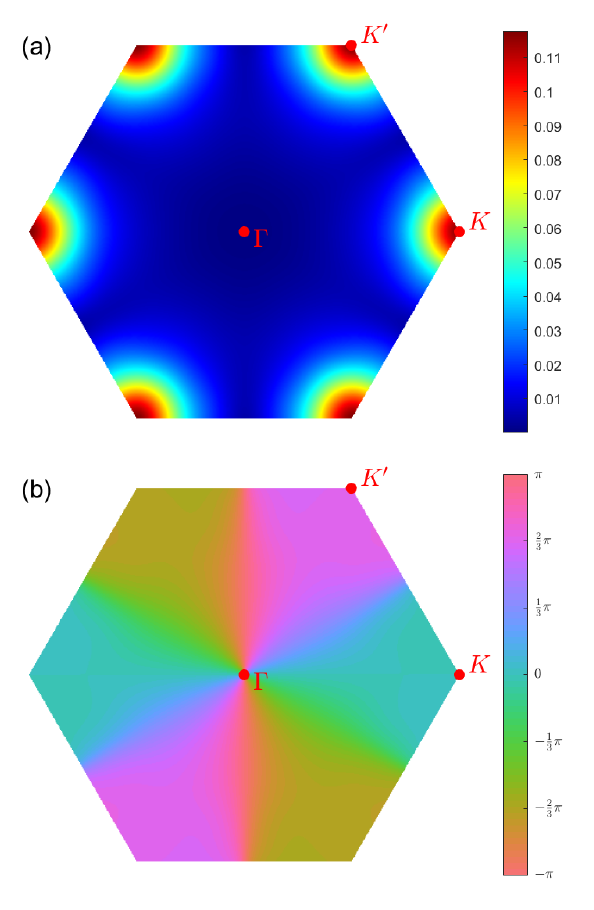}
\caption{\label{fig:d+id} Calculated (a) the absolute part $|\Delta_{\mathbf{k}}|$ and (b) the angle part $\theta_{\mathbf{k}}=Argz(\Delta_{\mathbf{k}})$ of the Cooper pair order parameter $\Delta_{\mathbf{k}}$, using $|\Psi_{+1}\rangle_{2\mathrm{h}}$ state ($L_z=+2$ variational ground state) in the 96-site sample.}
\end{figure}

Similarly, we can examine the Cooper pair order parameter in the momentum space:
\begin{equation}
\Delta_{\mathbf{k}}=_{2\mathrm{h}}\langle\Psi_{m}|c_{\mathbf{k}\uparrow}c_{-\mathbf{k}\downarrow}-c_{\mathbf{k}\downarrow}c_{-\mathbf{k}\uparrow}|\phi_0\rangle, \label{eq:Delta_k}
\end{equation}
in which $c_{\mathbf{k}\sigma}\equiv \frac{1}{\sqrt{2}}(c_{A\mathbf{k}\sigma}+c_{B\mathbf{k}\sigma})$. By decomposing $\Delta_{\mathbf{k}}$ into absolute and phase parts: $\Delta_{\mathbf{k}}=|\Delta_{\mathbf{k}}|e^{i\theta_{\mathbf{k}}}$, the calculated $|\Delta_{\mathbf{k}}|$ and $\theta_{\mathbf{k}}$ are shown in Fig.~\ref{fig:d+id}(a) and Fig.~\ref{fig:d+id}(b), respectively, for $|\Psi_{+1}\rangle_{2\mathrm{h}}$ in the 96-site sample. Here we fix $\Delta_{\mathbf{K^0}}$ to be a real number at $\mathbf{K^0}=(\frac{4\sqrt{3}\pi}{9},0)$ as the $K$ point in the Brillouin zone.

Fig.~\ref{fig:d+id} shows that the absolute part $|\Delta_{\mathbf{k}}|$ peaks at the $K$, $K^{\prime}$ points at the Brillouin zone boundary, which should be contributed from the pairing of the quasiparticle component peaked in Fig.~\ref{fig:nkh_96}. Then $\theta_{\mathbf{k}}$ increases from $0$ to $4\pi$ when $\mathbf{k}$ is rotated counterclockwise by 360 degrees from the $K$ direction. Two nodal lines exist in $\mathrm{Re}(\Delta_{\mathbf{k}})$ and $\mathrm{Im}(\Delta_{\mathbf{k}})$, which is the feature of a $d$ wave pairing to imply the $d+id$ pairing symmetry. Similarly, a $d-id$ pairing symmetry is found for the ansatz state $|\Psi_{-1}\rangle_{2\mathrm{h}}$.


\subsection{Pairing symmetry dichotomy: an $s$-wave pairing between ``twisted holes''}

\begin{figure}[h]
\includegraphics[width=0.4\textwidth]{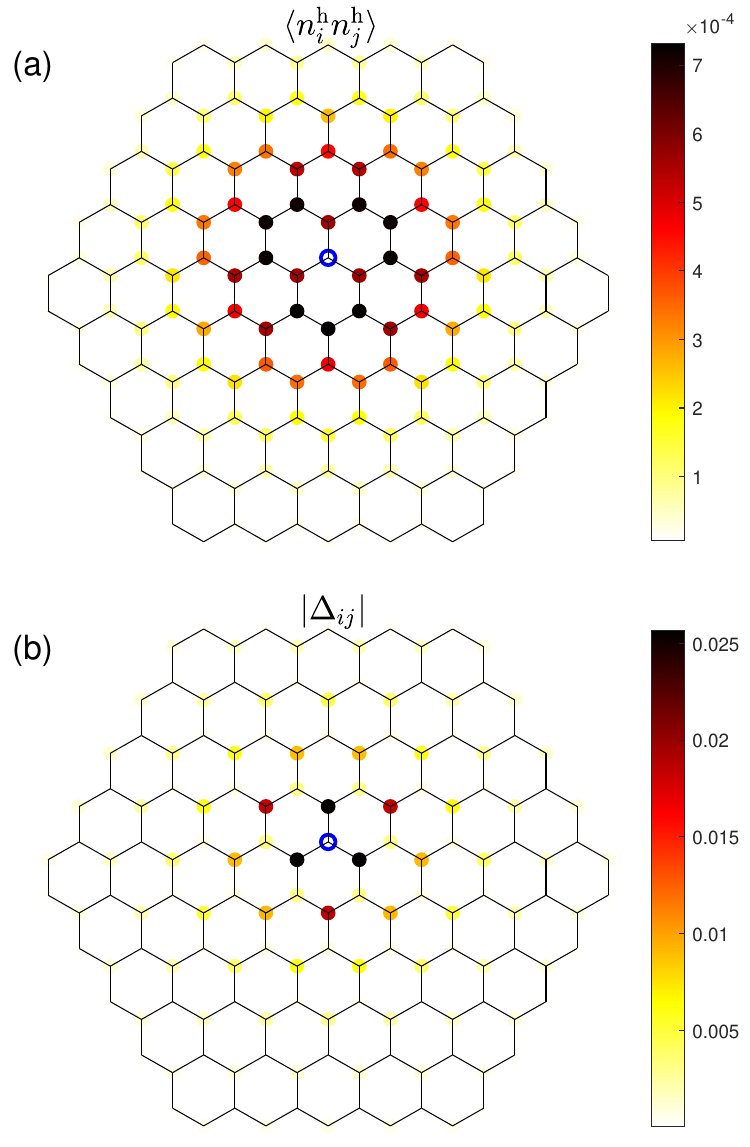}
\caption{\label{fig:dichotomy}(a) hole-hole density correlator $\langle n_i^h n_j^h\rangle$ for $L_z=2$ variational ground state in the 150-sites sample. One hole is fixed at the site $i$ with blue circle near the center of the sample. (b) The absolute value of Cooper pair order parameter $|\Delta_{ij}|$ for $L_z=2$ variational ground state in the 150-sites sample. Site $i$ is fixed near the center of the sample, denoted by the empty site with blue circle.}
\end{figure}

One may further study the hole-hole density correlator $\langle n_i^{\mathrm{h}} n_j^{\mathrm{h}}\rangle$ with $n_i^{\mathrm{h}}$ denoting the hole density at site $i$. By fixing the hole site $i$ in a 150-site sample using the ansatz state $|\Psi_{+1}\rangle_{2\mathrm{h}}$, the spatial pattern of $\langle n_i^{\mathrm{h}} n_j^{\mathrm{h}}\rangle$ is given in Fig.~\ref{fig:dichotomy}(a), which shows a pairing amplitude nearly isotropic and decays quickly as the distance $|\mathbf{r_i}-\mathbf{r_j}|$ increases. Here the $\langle n_i^{\mathrm{h}} n_j^{\mathrm{h}}\rangle$ reaches the largest value when the distance between the two holes is $2$. And the pairing amplitudes of distance $\sqrt{3}$ are very close to the largest value. 

On the other hand, Fig.~\ref{fig:dichotomy}(b) shows that the pattern of the real space Cooper pair order parameter in Eq.~\ref{eq:Delta_ij} is distinct from the above hole density distribution. When site $i$ and $j$ belong to different sublattices and $|\mathbf{r_i}-\mathbf{r_j}|$ is small ($\leq 2.65a_0$), the value of $|\Delta_{ij}|$ is relatively large ($\geq 0.009$). But at $|\mathbf{r_i}-\mathbf{r_j}|=1$, $|\Delta_{ij}|$ reach the largest whereas at $|\mathbf{r_i}-\mathbf{r_j}|=\sqrt{3}$, $|\Delta_{ij}|$ becomes very small ($\approx 0.003$).

To resolve this puzzle, we note that $|g(i,j)|$ in Eq.~(\ref{eq:twohole}) represents the pairing amplitude of the ``twisted holes'' described by $\tilde{c}_i \equiv c_ie^{\pm i\hat{\Omega}_i}$. Since $\langle n_i^{\mathrm{h}} n_j^{\mathrm{h}}\rangle\propto |g(i,j)|^2$, the isotropic density structure of the two holes is thus related to the $s$-wave pairing of the ``twisted holes''. Here, the $s$-wave pairing symmetry may be directly seen from the quantum numbers of the ground states. Consider the angular momentum $L_z=2$ for the two-hole ground state $|\Psi_{+1}\rangle_{2\mathrm{h}}$. It can be regarded as the direct summation of $L_z$ of two single-hole ground states, i.e. $(-2)+(-2)\equiv 2 (\mod 6)$, which means zero relative angular momentum for the two twisted holes, in consistency with the $s$-wave pairing. 

Therefore, the two-hole ground state has a pairing symmetry dichotomy: a $d+id$ pairing symmetry for the bare holes, $c_i$'s, emerges from the $s$-wave pairing in terms of twisted holes, $\tilde{c}_i$'s. The phase-shift factor $e^{\pm i\hat{\Omega}_i}$ acts a key role in such a dichotomy. And this pairing symmetry dichotomy has been previously revealed for the square lattice \cite{PhysRevX.12.011062}, which is thus a common characteristic of a bipartite $t$-$J$ model.

\subsection{Pairing energy and pairing size}

The pairing (binding) energy of the two holes may be defined by:
\begin{equation}
E_{\mathrm{pair}}=E_{2\mathrm{h}}+E_{0\mathrm{h}}-2E_{1\mathrm{h}},\label{eq:bindingE}
\end{equation}
where $E_{0\mathrm{h}}$, $E_{1\mathrm{h}}$, $E_{2\mathrm{h}}$ represent the half-filling, one-hole doped, two-hole doped ground state energy, respectively. 

\begin{figure}[h]
\includegraphics[width=0.4\textwidth]{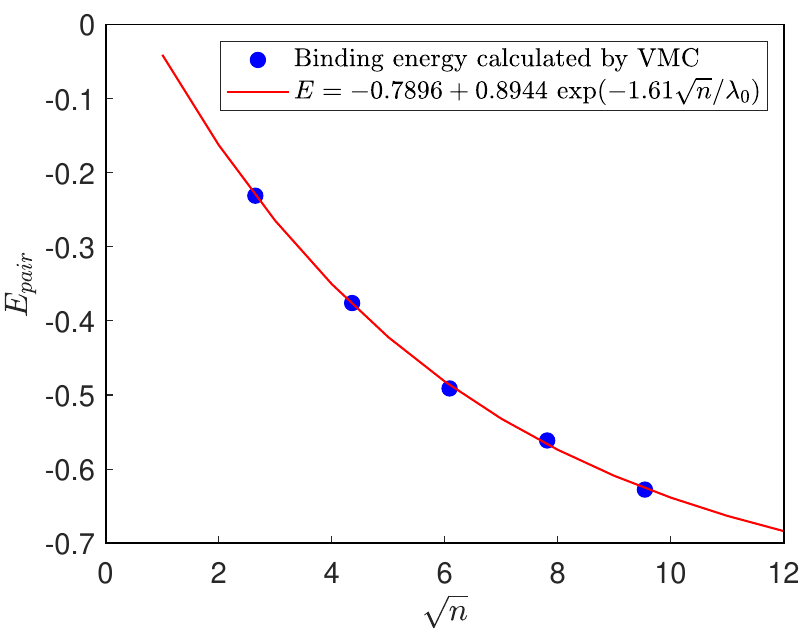}
\caption{\label{fig:binding_scaling}The binding energy $E_{pair}$ as the function of the $\sqrt{n}$ where $n$ is the total plaquette number of several samples. The blue dots represent the VMC calculation results. The red line denotes the fitting curve.}
\end{figure}

The scaling behavior of binding energies $E_{\mathrm{pair}}$ for systems with different sizes are shown in Fig.~\ref{fig:binding_scaling}. The area of $n$ plaquettes is given by $\frac{3\sqrt{3}}{2}n$, which defines a typical length scale $\lambda=\sqrt{\frac{3\sqrt{3}}{2}} \sqrt{n}\approx 1.61\sqrt{n}$. The pairing energy will saturate to a finite value $-0.7896J$ at large $n$. And the area ``occupied'' by the two holes is about $\lambda_0=9.05$ such that the pairing size is $4.53$, which is approximately the size of $\approx 4.017$ inferred from $|g(i,j)|$ as a function of $|\mathbf{r_i}-\mathbf{r_j}|$.  

\section{Discussion and conclusion}\label{sec:4}

In this work, we have examined the ground-state properties of the single-hole-doped and two-hole-doped $t$-$J$ model on honeycomb lattice based on the VMC study. For the simplest models of doped Mott insulators, a microscopic understanding of how the doped holes can propagate and form pairing in a quantum AFM spin background can provide valuable insight into the nature of doping Mott insulators.    

For the single-hole doping case, a nontrivial angular momentum $L_z=\pm 2$ is identified with a given total $S_z=\pm 1/2$ in the variational wave function, which is in agreement with the ED result. It means that the orbital part plays an important role in the profile of a single-hole composite. Indeed, a corresponding chiral spin (hole) current is found surrounding the doped hole (spin-1/2) mutually in the ground-state wave function Ansatz. 
Besides the overall good agreement with the finite-size ED calculation, including the total energy, the single-particle spectral weight $Z_{\mathbf{k}}$ is further shown to vanish in the large-sample limit. The present VMC study clearly shows that locally the doped hole has to ``twist'' the spins in the background into a vortex-like structure in order to facilitate its own hopping due to the underlying phase-string effect. Such a vortex-like ``twist'' in the spin background is manifested by an explicit angular momentum $L_z=\pm 2$ as contributed by the mutual circulation between a net spin-1/2 and the hole as illustrated in Fig.~\ref{fig:mutual}. It is noted that although the rest spin background does not directly contribute to $L_z$ as the rest of spins are all in RVB pairing, albeit long-ranged ones in an AFM state, they do play a crucial role in lowering the kinetic energy of the hole composite. 

For the two-hole doping case, a significant pairing between the holes is found, with the pairing symmetry in the Cooper channel to be purely $d+id$ form. However, based on the coordinates of the ``twisted'' holes mentioned above, an $s$-wave pairing is identified instead. Note that each single hole here has an angular momentum of $L_z=\pm 2$ and the two-hole ground state has the total spin zero (singlet) and angular momentum $L_z=\pm 4= \mp 2 \mod 6$, consistent with the $s$-wave pairing of the twisted holes. Namely a coherent $d+id$ Cooper pairing is just an inherited component of a two-hole state with a hidden $s$-wave pairing of the ``twisted'' holes. Without the latter, the Cooper pairing cannot exist alone as demonstrated by the variational wave function. Such a dichotomy in the symmetry of the hole pairing has been previously found in the square lattice case \cite{PhysRevX.12.011062}. Here the binding energy $0.79$$J$ is less than $1.97$$J$ in the square lattice case and the length scale of the pair size is about twice larger, which indicates that the pairing strength is weaker in the honeycomb lattice geometry due to the distinct geometries. Nonetheless, the underlying pairing mechanisms are the same as shown by the forms of the two-hole wave function Ansatz. Namely, as the single holes are strongly frustrated in the quantum spin background due to the phase-string effect, the frustration can get substantially released for two doped holes by forming a tightly bound pair. 

Here the pairing is irrelevant to the AFM long-range ordering of spins in the backdrop. Given such a small-size pairing as a building block, at least in the dilute finite doping regime, more holes are expected to form a ground state composed of similar building blocks to the leading order of approximation. Similar to the case on square lattice \cite{PhysRevX.12.011062}, the two-hole ground state in Eq. (\ref{eq:twohole}) ($m=\pm 1$) may be reexpressed in the following form: $|\Psi_{\mathrm{G}}\rangle_{2\mathrm{h}}=\hat{\mathcal{D}}|\mathrm{RVB}\rangle$ where $\hat{\mathcal{D}}$ denotes the creation of two ``twisted holes'' on top of a half-filling spin background $|\mathrm{RVB}\rangle$. Here 
\begin{equation}
\hat{\mathcal{D}}=\sum_{ij}g(i,j)\tilde {c}_{i\uparrow}\tilde {c}_{j\downarrow}
\end{equation}
with $g(i,j)$ is the $s$-wave pairing amplitude for the ``twisted holes'' obtained by the VMC scheme in this work, and $\tilde {c}_{i\sigma}\equiv {c}_{i\sigma}e^{\pm i\hat{\Omega}_i}$. Compared to the wave function in Eq. (\ref{eq:twohole}), the ``twisted'' spin background $|\mathrm{RVB}\rangle$ should be related to the original AFM state $|\phi_0\rangle$ via a transformation $e^{\mp i 2\hat{\Omega}_v}$ with $v=i$ or $j$, or simply smeared in the local two-hole region, which can be variationally determined in the VMC scheme \cite{PhysRevX.12.011062}. The implication is that in order for $\hat{\mathcal{D}}$ to describe an identical charge pair as a mobile object in total, the background must be simultaneously modified into $|\mathrm{RVB}\rangle$. Then a straightforward generalization to finite doping will give rise to the following superconducting ground-state \emph{Ansatz} originally proposed in Ref. \cite {Weng_2011}: 
\begin{equation}
    |\Psi_{\mathrm{G}}\rangle=e^{\hat{\mathcal{D}}}|\mathrm{RVB}\rangle,
\end{equation}
where a short-range AFM or RVB state $|\mathrm{RVB}\rangle$ will emerge self-consistently at finite doping \cite {Weng_2011,Weng2011}. How to realize such a superconducting state with a local AFM background from the present two-hole study, where the two-hole pair may be self-trapped in the original AFM spin background $|\phi_0\rangle$, will be an interesting problem to be explored elsewhere.

\begin{acknowledgments}
Useful discussions with Z. C. Gu, Z. L. Wang and Z. J. Song are acknowledged. This work is supported by MOST of China (Grant No. 2021YFA1402101) and NSF of China (Grant No. 12347107).

\end{acknowledgments}

\appendix

\section{\label{app:secA}Variational method used at half-filling}
At half-filling case, the $t$-$J$ model reduces to the Heisenberg model, whose ground state is assumed to be RVB state by Anderson \cite{Liang1988}:
\begin{equation}
|\phi_0\rangle=\sum_{v}\omega_{v}|v\rangle ,\label{eq:A1} 
\end{equation}
where the valence bond (VB) state $|v\rangle$ consists of spin-singlet pairs $(i,j)\equiv\frac{1}{\sqrt{2}}(|\uparrow_{i}\downarrow_{j}\rangle-|\downarrow_{i}\uparrow_{j}\rangle)$ on any two sites $i$ and $j$ belonging to different sublattices.

The amplitude $\omega_{\upsilon}$ of each VB configuration $v$ can be further factorized as $\omega_{\upsilon}=\prod_{(i,j)\in\upsilon}h_{ij}$, where $h_{ij}$ is the positive ``bond amplitude'' related with the spin-singlet pair connecting to site $i$ and $j$. For the Monte Carlo method used in this paper, we use $h_{ij}$ as our transition probability in the states updation procedure when sampling the VB states, following Ref. \cite{PhysRevB.82.024407}. The variational ground state $|\phi_0\rangle$ can be obtained by adjusting the bond amplitude $h_{ij}$ to minimize the energy. 

At half-filling, the ground state energy (having subtracted $-\frac{1}{4}n_in_j$ term in Hamiltonian) 
calculated by the VMC method above for the 24-site system shown in Fig.~\ref{fig:Lattice_24} is $-0.3570 J$ per bond, which is very close to the exact value $-0.3586 J$ per bond calculated by ED. 
We also note that similar to the square lattice case \cite{Liang1988}, the optimized wave function by VMC here also has the AFLRO, which can be seen by the spin-spin correlation $|\langle \mathbf{S}_{x_1}\cdot\mathbf{S}_{x_2} \rangle| \rightarrow 0.0972$ as $|x_1-x_2|\rightarrow \infty$.


\section{\label{app:secB}Symmetry analysis of the two-hole $Ansatz$ state}

In Sec. \ref{sec:3}, we show that the two-hole wave function can be viewed as an $s$ wave pairing of two ``twisted'' holes, i.e. two holes with spin-current vortex and antivortex, which leads to the $L_z=\pm 2$ degeneracy of the ground state. Here we show that the two degenerate ground states can be connected through time reversal transformation and are invariant under other symmetries of the Hamiltonian.

We first check how the ansatz states in Eq.~(\ref{eq:twohole}) change under time reversal operation. The definition of the time reversal operator $\hat{T}$ is:
\begin{equation}
\hat{T}c_{i\sigma}\hat{T}^{-1}=\sigma c_{i\bar{\sigma}}.
\end{equation}

The anti-unitary $\hat{T}$ acts on the phase shift operator $e^{im\hat{\Omega}_i}$ as:
\begin{equation}
\begin{split}
\hat{T}e^{im\hat{\Omega}_i}\hat{T}^{-1}&=e^{-im\sum_{l(\neq i)}\theta_i(l)n_{l\uparrow}}\\
&=e^{-im\sum_{l(\neq i)}\theta_i(l)}e^{+im\hat{\Omega}_i}.
\end{split}
\end{equation}
In the second equality above, we use the constrain $n_{l\uparrow}=1-n_{l\downarrow}$ at half-filling state $|\phi_0\rangle$.

Under time reversal transformation, 
\begin{equation}
\begin{split}
\hat{T}|\Psi_{m}\rangle_{2\mathrm{h}}=&\sum_{i j} e^{-im[\sum_{l(\neq i)}\theta_i(l)-\sum_{l(\neq j)}\theta_j(l)]}\\
&\times g_m^*(i, j) c_{j \uparrow}c_{i \downarrow}e^{im(\hat{\Omega}_i- \hat{\Omega}_j)}|\phi_0\rangle\\
=&\sum_{i j} e^{im[\sum_{l(\neq i)}\theta_i(l)-\sum_{l(\neq j)}\theta_j(l)]}\\
&\times g_m^*(j, i) c_{i \uparrow}c_{j \downarrow}e^{-im(\hat{\Omega}_i- \hat{\Omega}_j)}|\phi_0\rangle\\
=&e^{i\theta_T}|\Psi_{-m}\rangle_{2\mathrm{h}},\label{eq:D3}
\end{split}
\end{equation}
where $e^{i\theta_T}$ is an arbitrary $U(1)$ phase. The last equality in Eq.~(\ref{eq:D3}) requires:
\begin{equation}
g^*_{m}(j,i)=g_{-m}(i,j)e^{i\theta_{T}}e^{-im[\sum_{l(\neq i)}\theta_i(l)-\sum_{l(\neq j)}\theta_j(l)]}.
\end{equation}

Eq.~(\ref{eq:D3}) shows that time reversal operation can change one ansatz state $|\Psi_{m}\rangle_{\mathrm{2h}}$ to the other ansatz state $|\Psi_{-m}\rangle_{\mathrm{2h}}$ with opposite $L_z$. Since hole currents will change direction under time reversal transformation, the chiralities of hole currents in two degenerate ground states are opposite. 

Next, we consider a spin-flip symmetry, which reverse the $z$ component spin of electrons: 
\begin{equation}
    \hat{F}c_{i\sigma}\hat{F}^{-1}= c_{i\bar{\sigma}}.
\end{equation}
\begin{equation}
\begin{split}
\hat{F}e^{im\hat{\Omega}_i}\hat{F}^{-1}&=e^{+im\sum_{l(\neq i)}\theta_i(l)n_{l\uparrow}}\\
&=e^{+im\sum_{l(\neq i)}\theta_i(l)}e^{-im\hat{\Omega}_i}.
\end{split}
\end{equation}

Under spin-flip transformation:
\begin{equation}
\begin{split}
\hat{F}|\Psi_{m}\rangle_{2\mathrm{h}}=&\sum_{i j} -e^{+im[\sum_{l(\neq i)}\theta_i(l)-\sum_{l(\neq j)}\theta_j(l)]}\\
&\times g_m(i, j) c_{j \uparrow}c_{i \downarrow}e^{-im(\hat{\Omega}_i- \hat{\Omega}_j)}|\phi_0\rangle\\
=&\sum_{i j} -e^{-im[\sum_{l(\neq i)}\theta_i(l)-\sum_{l(\neq j)}\theta_j(l)]}\\
&\times g_m(j, i) c_{i \uparrow}c_{j \downarrow}e^{+im(\hat{\Omega}_i- \hat{\Omega}_j)}|\phi_0\rangle\\
=&e^{i\theta_F}|\Psi_{m}\rangle_{2\mathrm{h}},\label{eq:D7}
\end{split}
\end{equation}
where $e^{i\theta_F}$ is an arbitrary $U(1)$ phase. The last equality in Eq.~(\ref{eq:D3}) requires:
\begin{equation}
g_{m}(j,i)=-g_{m}(i,j)e^{i\theta_{F}}e^{+im[\sum_{l(\neq i)}\theta_i(l)-\sum_{l(\neq j)}\theta_j(l)]}.
\end{equation}

Therefore, the ansatz state $|\Psi_{m}\rangle_{\mathrm{2h}}$ remains unchanged un to a trivial phase under spin-flip transformation. Since spin currents will change direction under spin-flip transformation, Eq.~(\ref{eq:D7}) guarantees there are no net spin currents in the two-hole ground state.

\section{\label{app:secC}Variational Monte Carlo method used in hole-doping case}

In this appendix, we give a brief description of the variational Monte Carlo procedure used to calculate physical quantities for both single-hole and two-hole doped cases. 

The expectation value of an operator $\hat{O}$ over the single-hole or two-hole variational wave function $|\Psi\rangle$ can be generally written as:
\begin{equation}
\langle\hat{O}\rangle\equiv\frac{\langle\Psi|\hat{O}|\Psi\rangle}{\langle\Psi|\Psi\rangle}=\frac{\langle\Psi|\hat{O}|\Psi\rangle}{\langle\phi_0|\phi_0\rangle}\frac{\langle\phi_0|\phi_0\rangle}{\langle\Psi|\Psi\rangle}.\label{eq:B1}
\end{equation}

By fixing the normalization condition as
\begin{equation}
\frac{\langle\Psi|\Psi\rangle}{\langle\phi_0|\phi_0\rangle}=1,~\label{eq:B2}
\end{equation}
we obtain the formula to calculate $\langle\hat{O}\rangle$.
For single-hole case:
\begin{widetext}
\begin{equation}
\langle\hat{O}\rangle=\sum_{i,j}\varphi_{\mathrm{h}}^{*(m)}(j)\varphi^{(m)}_{\mathrm{h}}(i)\left[\sum_{v^{\prime},v}P\left(v^{\prime},v\right)\frac{\left\langle v^{\prime}|e^{+im\hat{\Omega}_{j}}c_{j}^{\dagger}\hat{O}c_{i}e^{-im\hat{\Omega}_{i}}|v\right\rangle}{\langle v^{\prime}\mid v\rangle}\right],\label{eq:B3}
\end{equation}
and for two-hole case:
\begin{equation}
\langle\hat{O}\rangle=\sum_{i,j,i',j'}g_m^{*}(i',j')g_m(i,j)\left[\sum_{v^{\prime},v}P\left(v^{\prime},v\right)\frac{\langle v^{\prime}|e^{-im(\hat{\Omega}_{i'}-\hat{\Omega}_{j'})}c_{j'\downarrow}^{\dagger}c_{i'\uparrow}^{\dagger}\hat{O}c_{i\uparrow}c_{j\downarrow}e^{+im(\hat{\Omega}_i-\hat{\Omega}_j)}|v\rangle}{\langle v^{\prime}|v\rangle}\right]. \label{eq:B4}
\end{equation}

where 
\begin{equation}
P\left(v^{\prime},v\right)\equiv\frac{w_{v^{\prime}}w_{v}\left<v^{\prime}|v\right>}{\left<\phi_0|\phi_0\right>}
\end{equation}
is the probability distribution of VB configurations, which satisfies the normalization condition:$\sum_{v,v'} P\left(v^{\prime},v\right)=1$. It can be used for Monte Carlo sampling as the half-filling case in Appendix~\ref{app:secA}.

For single-hole-doping case, we can define matrix element:
\begin{equation}
\mathbf{O}^{j}_{\,i}=\sum_{v^{\prime},v}P\left(v^{\prime},v\right)\frac{\langle v^{\prime}|e^{+im\hat{\Omega}_{j}}c_{j\uparrow}^{\dagger}\hat{O}c_{i\uparrow}e^{-im\hat{\Omega}_{i}}|v\rangle}{\langle v^{\prime}|v\rangle}. \label{eq:B7}
\end{equation}

Similarly, for two-hole-doping case, we can define matrix element:
\begin{equation}
\mathbf{O}^{i'j'}_{\,ij}=\sum_{v^{\prime},v}P\left(v^{\prime},v\right)\frac{\langle v^{\prime}|e^{-im(\hat{\Omega}_{i'}-\hat{\Omega}_{j'})}c_{j'\downarrow}^{\dagger}c_{i'\uparrow}^{\dagger}\hat{O}c_{i\uparrow}c_{j\downarrow}e^{+im(\hat{\Omega}_i-\hat{\Omega}_j)}|v\rangle}{\langle v^{\prime}|v\rangle}. \label{eq:B8}
\end{equation}
\end{widetext}

If we view the variational parameter $\varphi^{(m)}_{\mathrm{h}}(i)$ or $g_m(i,j)$  
as the column vector $\mathbf{x}$, then $\langle \hat{O} \rangle$ in Eq.~(\ref{eq:B3}) and Eq.~(\ref{eq:B4}) can be expressed as a quadratic form: 
\begin{equation}
\langle\hat{O}\rangle=\mathbf{x}^{\dagger}\mathbf{O}\mathbf{x}. \label{eq:eigenvector}
\end{equation}

Specifically, Eq.~(\ref{eq:B2}) can be seen as the special case where $\hat{O}=\hat{I}$ in Eq.~(\ref{eq:B3}) or Eq.~(\ref{eq:B4}), which are denoted as:
\begin{equation}
1=\langle\hat{I}\rangle=\mathbf{x}^\dagger\mathbf{A}\mathbf{x}.~\label{eq:B9}
\end{equation}

Eq.~\eqref{eq:B9} imposes the normalization condition to the variational wave function.
By optimizing the total energy $\langle H_{t\text{-}J} \rangle$ within the constrain Eq.~(\ref{eq:B9}), 
the procedure to find the ground state turns out to be a generalized eigenvalue problem:
\begin{equation}
\mathbf{H}_{t\text{-}J}\mathbf{x}=E_{t\text{-}J}\mathbf{A}\mathbf{x}.~ \label{eq:B10}
\end{equation}

The wave function of variational ground state is the generalized eigenvector that corresponds to the minimum energy $E_{t\text{-}J}$. Once the ground state wave function is solved, the expectation of any observable can be calculated using Eq.~(\ref{eq:eigenvector}).

Following the method in Ref
.~\cite{PhysRevX.12.011062}, we transform the operators in Eq.~(\ref{eq:B3}) and Eq.~(\ref{eq:B4}) to the form below to calculate the matrix elements:
\begin{equation}
\frac{\langle v'|\hat{\Lambda}^{-m}n_{k_1\sigma_1}n_{k_2\sigma_2}\cdots n_{k_n\sigma_n}S_{l_1}^{\sigma_1}S_{l_2}^{\sigma_2}\cdots S_{l_s}^{\sigma_s}\hat{\Lambda}^{m}|v\rangle}{\langle v'|v\rangle},\label{eq:B11}
\end{equation}
where $\hat{\Lambda}^{m}=e^{-im\hat{\Omega}_{i}}$ for single-hole-doping case and $\hat{\Lambda}^{m}=e^{+im(\hat{\Omega}_i-\hat{\Omega}_j)}\equiv \hat{\Lambda}_{ij}^{m}$ for two-hole-doping case.

We consider the loop configurations in the transposition-graph covers $(v,v^{\prime})$ and check the compatibility of the operator $n_{k_n\sigma_n}$ or $S_{l_s}^{\sigma_s}$ with each loop when calculating the phase factor $\hat{\Lambda}^{m}$. 

The expressions of the matrix elements of the operators used in the main text are listed below.

\subsection{Explicit expressions of matrix elements used in single-hole-doping case}

 (1) The normalization matrix $\mathbf{A}$ defined in Eq.~(\ref{eq:B9}) is proportional to identity matrix:
 \begin{equation}
\mathbf{A}^{i'}_{\,i}=\frac{1}{2}\delta_{i,i'}.
 \end{equation}

(2) The matrix elements of the hopping term $H_t$ in Eq.~(\ref{eq:Ht}) are:
\begin{equation}\begin{aligned}(\mathbf{H}_{t})_{i}^{i^{\prime}}&=t\sum_{v^{\prime},v}P(v^{\prime},v)\frac{1}{\langle v^{\prime}|v\rangle}\delta_{i,\text{NN}(i^{\prime})}\\&\times\langle v^{\prime}|e^{+im\hat{\Omega}_{i^{\prime}}}(n_{i\uparrow}n_{i^{\prime}\uparrow}+S_{i}^{-}S_{i^{\prime}}^{+})e^{-im\hat{\Omega}_{i}}|v\rangle,\end{aligned}
\end{equation}
where $\delta_{i,\text{NN}(i^{\prime})}=1$ when one of the nearest neighbors of site $i'$ is site $i$ and $\delta_{i,\text{NN}(i^{\prime})}=0$ otherwise.

The hole current $J^h_{ij}$ in Eq.~(\ref{eq:J_h}) can be obtained by taking the imaginary part of the matrix $(\mathbf{H}_{t})_{i}^{i^{\prime}}$:
\begin{equation}
J^h_{ij}=2\mathrm{Im}(\varphi^{*(m)}_{\mathrm{h}}(j)(\mathbf{H}_{t})_{i}^{j}\varphi^{(m)}_{\mathrm{h}}(i)).
\end{equation}

(3) The matrix elements of the superexchange term $H_J$ in Eq.~(\ref{eq:HJ}) are:
\begin{equation}
\begin{aligned}(\mathbf{H}_{J})_{i}^{i'}&=\frac{J}{2}\sum_{v',v,\langle kl\rangle(\neq i)}P(v',v)\frac{1}{\langle v'|v\rangle}\delta_{ii'}\langle v'|e^{+im\hat{\Omega}_{i'}}n_{i\uparrow}\\&\times(S_k^{+}S_l^{-}+S_k^{-}S_l^{+}-n_{k\uparrow}n_{l\downarrow}-n_{k\downarrow}n_{l\uparrow})e^{-im\hat{\Omega}_i}|v\rangle.
\end{aligned}
\end{equation}

(4) The matrix elements of the neutral spin current $J^s_{ij}$ in Eq.~(\ref{eq:J_s}) are:
\begin{equation}
\begin{aligned}(\mathbf{J}^s_{kl})_{i}^{i'}&=i\frac J2(1-\delta_{ik})(1-\delta_{il})\sum_{v',v,\langle kl\rangle(\neq i)}P(v',v)\frac{1}{\langle v'|v\rangle}\\&\delta_{ii'}\langle v'|e^{+im\hat{\Omega}_{i'}}n_{i\uparrow}(S_k^{+}S_l^{-}-S_k^{-}S_l^{+})e^{-im\hat{\Omega}_i}|v\rangle.
\end{aligned}
\end{equation}

(5) The calculation procedures of spectral weight $Z_{\mathbf{k}1}$ or $Z_{\mathbf{k}2}$ in Eq.~(\ref{eq:Zkupper}) and Eq.~(\ref{eq:Zklower}) are different from above. For simplicity, we first consider the overlap $\langle\phi_0|c^{\dagger}_{\eta\mathbf{k}\uparrow}|\Psi_m\rangle_{1\mathrm{h}}$ where $\eta=A,B$ denote for A and B sublattice. We normalize it as:
\begin{equation}
\begin{aligned}
&\frac{\langle\phi_0|c^{\dagger}_{\eta\mathbf{k}\uparrow}|\Psi_m\rangle_{1\mathrm{h}}}{\sqrt{\langle\phi_0|\phi_0\rangle_{[1h]}\langle\Psi_\mathrm{m}|\Psi_\mathrm{m}\rangle_{1\mathrm{h}}}}\\
=&\sum_{i}\varphi_{\mathrm{h}}^{m}(i)\frac{\langle\phi_{0}|c_{\eta\mathrm{k}\uparrow}^{\dagger}c_{i\uparrow}e^{-im\hat{\Omega}_{i}}|\phi_{0}\rangle}{\langle\phi_{0}|\phi_{0}\rangle}\\
=&\sum_{i\in \eta}\frac{1}{\sqrt{N}}e^{i\mathbf{k}\cdot\mathbf{x}_i}\varphi_{\mathrm{h}}^{m}(i)\sum_{v^{\prime},v}P(v^{\prime},v)\frac{\langle v^{\prime}|n_{i\uparrow}e^{-im\hat{\Omega}_{i}}|v\rangle}{\langle v^{\prime}|v\rangle}. \label{eq:C17}
\end{aligned}
\end{equation}

Since $\alpha_{\mathbf{k}}$ and $\beta_{\mathbf{k}}$ can be written as the linear combination of $c_{A\mathbf{k}\sigma}$ and $c_{B\mathbf{k}\sigma}$, $Z_{\mathbf{k}1}$ or $Z_{\mathbf{k}2}$ can be obtained from the results in Eq.~(\ref{eq:C17}).

\subsection{Explicit expressions of matrix elements used in two-hole-doping case}

(1) The normalization matrix $\mathbf{A}$ defined in Eq.~(\ref{eq:B9}) is:
\begin{widetext}
\begin{equation}
    \mathbf{A}^{i'j'}_{ij}= \sum_{v',v}P(v',v)\frac{1}{\langle v'|v\rangle}
    \left[\delta_{ii'}\delta_{jj'}\langle v'|\hat{\Lambda}^{-m}_{ij}n_{i\uparrow}n_{j\downarrow}\hat{\Lambda}^{m}_{ij}|v\rangle
    -\delta_{ij'}\delta_{ji'}\langle v'|\hat{\Lambda}^{-m}_{ji}S_i^-S_j^+\hat{\Lambda}^{m}_{ij}|v\rangle \right] ~.
    \label{eqn:measurenorm}
\end{equation}

(2) The matrix elements of the hopping term $H_t$ in Eq.~(\ref{eq:Ht}) are:
\begin{equation}
    \begin{aligned}
        (\mathbf{H}_{t})_{ij}^{i'j'} =& t \sum_{v',v}P(v',v)\frac{1}{\langle v'|v\rangle}
        \left[\delta_{ii'}\delta_{j,\mathrm{NN}(j')}\langle v'|\hat{\Lambda}^{-m}_{ij'}n_{i\uparrow}(n_{j\downarrow}n_{j'\downarrow}
        +S_j^+S_{j'}^-)\hat{\Lambda}^{m}_{ij}|v\rangle\right.\\
        &\left. +\delta_{jj'}\delta_{i,\mathrm{NN}(i')}\langle v'|\hat{\Lambda}^{-m}_{i'j} n_{j\downarrow} (n_{i\uparrow}n_{i'\uparrow}
        +S^-_iS^+_{i'})\hat{\Lambda}^{m}_{ij}|v\rangle \right.
        \left. -\delta_{ji'}\delta_{i,\mathrm{NN}(j')}\langle v'|\hat{\Lambda}^{-m}_{j'i}S^-_{i}(S^+_{j}n_{j'\uparrow}+n_{j\downarrow}S^
        +_{j'})\hat{\Lambda}^{m}_{ij} |v\rangle \right.\\
        &\left. -\delta_{ij'}\delta_{j,\mathrm{NN}(i')}\langle v'|\hat{\Lambda}^{-m}_{ji'}S^-_{j}(S^+_{i}n_{i'\uparrow}
        +n_{i\downarrow}S^+_{i'})\hat{\Lambda}^{m}_{ij} |v\rangle \right]~,\\
    \end{aligned}
    \label{eqn:measurehopping}
\end{equation}

(3) The matrix elements of the superexchange term $H_J$ in Eq.~(\ref{eq:HJ}) are:
\begin{equation}
    \begin{aligned}
        (\mathbf{H}_{J})_{ij}^{i'j'} =& \frac{J}{2} \sum_{v',v}P(v',v)\frac{1}{\langle v'|v\rangle}\sum_{\langle kl\rangle(\neq i,j)}
        \left[\delta_{ii'}\delta_{jj'}\langle v'|\hat{\Lambda}^{-m}_{ij}n_{i\uparrow}n_{j\downarrow}
        (S_k^+S_l^-+S_k^-S_l^+-n_{k\uparrow}n_{l\downarrow}-n_{k\downarrow}n_{l\uparrow})\hat{\Lambda}^{m}_{ij}|v\rangle \right.\\ 
        &\left.+ \delta_{ij'}\delta_{ji'}\langle v'|\hat{\Lambda}^{-m}_{ji}S_i^-S_j^+
        (n_{k\uparrow}n_{l\downarrow}+n_{k\downarrow}n_{l\uparrow}-S_k^+S_l^--S_k^-S_l^+)\hat{\Lambda}^{m}_{ij}|v\rangle \right]~.
    \end{aligned}
    \label{eqn:measureAF}
\end{equation}

(4) The calculation procedures of Cooper pair order parameter $\Delta_{\mathbf{k}}$ in Eq.~(\ref{eq:Delta_k}) are different from above. We first normalize it as: 
\begin{equation}
    \begin{aligned}
        \Delta_{\mathbf{k}} =& \frac{_{\mathrm{2h}}{\langle\Psi_m|\hat{\Delta}_{\mathbf{k}}^s|\phi_0\rangle}{}}
        {\sqrt{_{\mathrm{2h}}{\langle\Psi_m|\Psi_m\rangle}{_{\mathrm{2h}}}\langle\phi_0|\phi_0\rangle}} 
        =\sum_{i'j'm}g^*_{m}(i',j') \frac{ \langle\phi_0|\hat{\Lambda}^{-m}_{i'j'}c^\dagger_{j'\downarrow}c^{\dagger}_{i'\uparrow}\hat{\Delta}_{\mathbf{k}}^s|\phi_0\rangle} {\langle\phi_0|\phi_0\rangle}  \\
        =&\sum_{i'j',ij}g^*_{m}(i',j')\frac{1}{N}e^{i\mathbf{k}\dot(\mathbf{r}_i - \mathbf{r}_j)}\sum_{v',v}P(v',v)
         \frac{ \langle v'|\hat{\Lambda}^{-m}_{i'j'}c^\dagger_{j'\downarrow}c^{\dagger}_{i'\uparrow}\hat{\Delta}_{ij}^s\hat{\Lambda}_{ij}^0|v\rangle} {\langle v'|v\rangle} ~,\\
    \end{aligned}
\end{equation}
where $\hat{\Delta}^s_{\mathbf{k}}\equiv c_{\mathbf{k}\uparrow} c_{-\mathbf{k} \downarrow}-c_{\mathbf{k} \uparrow} c_{-\mathbf{k} \downarrow}$ and $\hat{\Delta}^s_{ij}\equiv c_{i \uparrow} c_{j \downarrow}-c_{j \uparrow} c_{i \downarrow}$ are pair operators in momentum space and real space, respectively. Viewing $\frac{1}{N}e^{i\mathbf{k}\dot(\mathbf{r}_i - \mathbf{r}_j)}$ as the column vector $g_0(i,j)$, $\Delta_{\mathbf{k}}$ can be written as the quadratic form in Eq.~(\ref{eq:B8}). The matrix elements of it are:
\begin{equation}
    \label{eqn:measureoverlap}
    \begin{aligned}
        (\mathbf{\Delta}_{\mathbf{k}})_{ij}^{i'j'} \equiv &\sum_{v',v}P(v',v)
        \frac{\langle v'|\hat{\Lambda}^{-m}_{i'j'}c^\dagger_{j'\downarrow}c^{\dagger}_{i'\uparrow}\hat{\Delta}_{ij}^s\hat{\Lambda}_{ij}^0|v\rangle}{\langle v'|v\rangle}\\
        =&\sum_{v',v}P(v',v)\frac{1}{\langle v'|v\rangle}
        \left[\delta_{ii'}\delta_{jj'}\langle v'|\hat{\Lambda}^{-m}_{ij}(n_{i\uparrow}n_{j\downarrow}-S_{i}^+S_{j}^-)|v\rangle 
        +\delta_{ij'}\delta_{ji'}\langle v'|\hat{\Lambda}^{-m}_{ji}(n_{i\downarrow}n_{j\uparrow}-S_{i}^-S_{j}^+)|v\rangle \right] ~.
    \end{aligned}
\end{equation}

Eq.~(\ref{eqn:measureoverlap}) implies that the expectation value of real-space Cooper pair order parameter in Eq.~(\ref{eq:Delta_ij}) can be calculated by:
\begin{equation}
    \label{eqn:measureoverlapij}
        \Delta_{ij}=\sum_{i'j'}g^*_{m}(i',j')\sum_{v',v}P(v',v)\frac{1}{\langle v'|v\rangle}
        \left[\delta_{ii'}\delta_{jj'}\langle v'|\hat{\Lambda}^{-m}_{ij}(n_{i\uparrow}n_{j\downarrow}-S_{i}^+S_{j}^-)|v\rangle 
        +\delta_{ij'}\delta_{ji'}\langle v'|\hat{\Lambda}^{-m}_{ji}(n_{i\downarrow}n_{j\uparrow}-S_{i}^-S_{j}^+)|v\rangle \right] ~.
\end{equation}

(5) Denote the matrix of the hole-hole correlator $n^{\mathrm{h}}_in^{\mathrm{h}}_j$ is $\mathbf{N}_{ij}^{\mathrm{h}}$. The expression of it is:
\begin{equation}
    (\mathbf{N}_{kl}^{\mathrm{h}})_{ij}^{i'j'} = (\delta_{ik}\delta_{jl}+\delta_{il}\delta_{jk})\sum_{v',v}P(v',v)\frac{1}{\langle v'|v\rangle}
    \left[\delta_{ii'}\delta_{jj'}\langle v'|\hat{\Lambda}^{-m}_{ij}n_{i\uparrow}n_{j\downarrow}\hat{\Lambda}^{m}_{ij}|v\rangle
    -\delta_{ij'}\delta_{ji'}\langle v'|\hat{\Lambda}^{-m}_{ji}S_i^-S_j^+\hat{\Lambda}^{m}_{ij}|v\rangle \right]~.
    \label{eqn:measurenihnjh}
\end{equation}

(6) The matrix elements of the neutral spin current $J^s_{ij}$ in Eq.~(\ref{eq:J_s}) are:
\begin{equation}
    \begin{aligned}
        (\mathbf{J}_{kl}^s)^{i'j'}_{ij} =& i\frac{J}{2} (1-\delta_{\langle ij\rangle,\langle kl\rangle})\sum_{v',v}P(v',v)\frac{1}{\langle v'|v\rangle}
        \left[\delta_{ii'}\delta_{jj'}\langle v'|\hat{\Lambda}^{-m}_{ij}n_{i\uparrow}n_{j\downarrow}
        (S_k^+S_l^--S_k^-S_l^+)\hat{\Lambda}^{m}_{ij}|v\rangle\right.\\ 
        &\left.- \delta_{ij'}\delta_{ji'}\langle v'|\hat{\Lambda}^{-m}_{ji}S_i^-S_j^+ (S_k^+S_l^--S_k^-S_l^+)\hat{\Lambda}^{m}_{ij}|v\rangle \right] ~,
    \end{aligned}
    \label{eqn:measureJs}
\end{equation}
where $\delta_{\langle ij\rangle,\langle kl\rangle} = 0$ when $i\neq k,l$ and $j\neq k,l$, and $\delta_{\langle ij\rangle,\langle kl\rangle} = 1$ otherwise. 

When the hole with spin $\uparrow$ is projected at site $i_0$ and the other hole with spin $\downarrow$ is projected at site $j_0$ (See Fig.~\ref{fig:Js_2h}), the matrix elements of neutral spin current are: 
\begin{equation}
(\mathbf{J}_{kl}^s)^{i'j'}_{ij} (\text{projected})= i\frac{J}{2} (1-\delta_{\langle ij\rangle,\langle kl\rangle})\sum_{v',v}P(v',v)\frac{1}{\langle v'|v\rangle}\delta_{ii'}\delta_{jj'}\delta_{ii_0}\delta_{jj_0}\langle v'|\hat{\Lambda}^{-m}_{ij}n_{i\uparrow}n_{j\downarrow}(S_k^+S_l^--S_k^-S_l^+)\hat{\Lambda}^{m}_{ij}|v\rangle .
    \label{eqn:measureJstwohole}
\end{equation}

\end{widetext}
\section{\label{app:secD}The longitudinal spin polaron effect incorporation}

In Sec. \ref{sec:2}, we show that the phase-shift operator $e^{-im\hat{\Omega}_{i}}$ in Eq.~(\ref{eq:singlehole}) leads to the \emph{transverse} spin current pattern around the hole, which plays the most essential role in facilitating the hole's hopping.

On the other hand, the conventional self-consistent Born approximation theory (SCBA) considers the renormalization of the hole's effective mass due to the scattering between holon and spin magnon excitations \cite{PhysRevB.2.1324,PhysRevLett.60.2793}, which is identified as the ``longitudinal spin polaron effect'' in this paper. 

This effect can also be incorporated here by making use of the Lanczos algorithm. The basic idea of Lanczos algorithm is to construct a new state by linearly combining variational ground state $|\Psi\rangle$ and $\hat{H}|\Psi\rangle$ to further lower the variational energy. Based on a similar idea, we put forward the ``improved" wave function $\emph{Ansatz}$ below:

\begin{equation}
\begin{split}
|\tilde{\Psi}_{m}\rangle_{1\mathrm{h}} = &\sum_{i}{\Big(\varphi^{(m)}_{1}(i)c_{i\uparrow}e^{-im\hat{\Omega}_{i}}+\sum_{j\in NN(i)}(\varphi^{(m)}_{2}(i,j)n_{i\uparrow}c_{j\uparrow}\Big. } \\
&{\Big. +\varphi^{(m)}_{3}(i,j)S^{-}_{i}c_{j\downarrow})e^{-im\hat{\Omega}_{j}}\Big)}|\phi_0\rangle,
\label{eq:lanczos}
\end{split}
\end{equation}
where the $\varphi^{(m)}_{1}(i)$, $\varphi^{(m)}_{2}(i,j)$ and $\varphi^{(m)}_{3}(i,j)$ are variational parameters to be optimized. 
The extra variational parts compared to $|\Psi_{m}\rangle_{\mathrm{1h}}$ in Eq.~(\ref{eq:singlehole})
can be seen as the longitudinal spin polaron effect to the spin background, which transforms the spin background $|\phi_0\rangle \rightarrow \hat{\Pi}_i|\phi_0\rangle$ as in Eq.~(\ref{eq:tildePsi}). 
Therefore, the definition of $\hat{\Pi}_i$ in Eq.~(\ref{eq:tildePsi}) is: 
\begin{equation}
\hat{\Pi}_i=1+\sum_{j\in\mathrm{NN}(i)}[a_1(i,j)n_{j\uparrow}+a_2(i,j)S_j^-],
\end{equation}
where $a_1(i,j)$ and $a_2(i,j)$ are variational parameters.



\bibliography{apssamp}

\end{document}